\documentclass[preprint,showpacs,amsmath,amssymb,prc,floatfix]{revtex4}
\usepackage{graphicx}

\begin{document}

\title{Hadron attenuation in deep inelastic lepton-nucleus scattering}
\author{T. Falter}
\email{Thomas.Falter@theo.physik.uni-giessen.de}
\author{W. Cassing}
\author{K. Gallmeister}
\author{U. Mosel}
\affiliation{Institut fuer Theoretische Physik, Universitaet
Giessen\\ D-35392 Giessen, Germany}
\date{\today}
\begin{abstract}
We present a detailed theoretical investigation of hadron attenuation in deep inelastic scattering (DIS) off complex nuclei in the kinematic regime of the HERMES experiment. The analysis is carried out in the framework of a probabilistic coupled-channel transport model based on the Boltzmann-Uehling-Uhlenbeck (BUU) equation, which allows for a treatment of the final-state interactions (FSI) beyond simple absorption mechanisms. Furthermore, our event-by-event simulations account for the kinematic cuts of the experiments as well as the geometrical acceptance of the detectors. We calculate the multiplicity ratios of charged hadrons for various nuclear targets relative to deuterium as a function of the photon energy $\nu$, the hadron energy fraction $z_h=E_h/\nu$ and the transverse momentum $p_T$. We also confront our model results on double-hadron attenuation with recent experimental data. Separately, we compare the attenuation of identified hadrons ($\pi^\pm$, $\pi^0$, $K^\pm$, $p$ and $\bar{p}$) on $^{20}$Ne and $^{84}$Kr targets with the data from the HERMES Collaboration and make predictions for a $^{131}$Xe target. At the end we turn towards hadron attenuation on $^{63}$Cu nuclei at EMC energies. Our studies demonstrate that (pre-)hadronic final-state interactions play a dominant role in the kinematic regime of the HERMES experiment while our present approach overestimates the attenuation at EMC energies.
\end{abstract}

\pacs{24.10.-i,25.30.-c,13.60.Le}
\maketitle

\section{Introduction}
\label{sec:intro} 
During the past years the HERMES collaboration has carried out a detailed experimental investigation of deep inelastic lepton scattering off complex nuclei \cite{HERMESDIS,HERMESDIS_new}. This has lead to numerous excellent data on the attenuation of high-energy hadrons in 'cold' nuclear matter which can be used to study the hadronization of particles created in (hard) collisions as proposed in Ref.~\cite{Kop}.

Besides being an interesting topic on its own the detailed understanding of the space-time picture of hadronization should also help to clarify to what extent the jet suppression observed in heavy-ion collisions at RHIC \cite{RHIC} is due to (pre-)hadronic final-state interactions (FSI) \cite{Gal03,Cass04} or partonic energy loss \cite{partonicRHIC}. In this respect the lepton scattering experiments also provide a benchmark for the interpretation of ultra-relativistic nucleus-nucleus collisions that aim at probing the transition from deconfined matter to a color neutral interacting hadron gas.

Contrary to nucleus-nucleus collisions, where the created 'fireball' is rapidly expanding, lepton-nucleus reactions provide rather clear geometrical constraints, which are well under control experimentally. In the one-photon exchange approximation the scattering of a charged lepton off a nucleon or nucleus can be decomposed into the emission and absorption of a virtual photon with energy $\nu$, virtuality $Q^2$ and polarization $\epsilon$. In such a high-energy photon-nucleon interaction the reaction products hadronize, due to confinement, long before they reach the detector. One can estimate the formation time of a hadron by the time that its partonic constituents need to travel a distance of the order of a hadronic radius. According to this estimate the formation time in the rest frame of the hadron is about 0.5--0.8 fm/$c$. Because of time dilatation, the formation length in the laboratory frame then can exceed nuclear radii at high energies. In lepton-nucleus interactions the reaction products can, therefore, interact with the surrounding nuclear medium during the formation time. Here the nuclear target can be viewed as a micro-laboratory that provides an intrinsic (variable) time scale due to the size of the target nuclei, which can be exploited to get information on the actual time scale of the hadronization itself.

The HERMES collaboration has measured the multiplicity ratios of pions, charged kaons, protons and antiprotons in deep inelastic positron scattering at 27.6 GeV off $^2$D, $^{14}$N, $^{20}$Ne and $^{84}$Kr targets. The observed attenuation of the hadron multiplicities in complex nuclei has lead to different interpretations. One possible explanation is that the quark, that was struck by the photon, propagates through the nucleus and undergoes multiple scattering \cite{Wang,Arl03}, thereby loosing energy by induced gluon radiation. The authors of Ref.~\cite{Acc02} attribute the observed attenuation to a combined effect of hadron absorption and a rescaling of the quark fragmentation function in nuclei. In Refs.~\cite{Fal03,Fal03b} we have shown that the attenuation of charged hadrons at HERMES energies can also be understood by final-state interactions (FSI) of color neutral prehadrons that are produced quite early after the photon-nucleon interaction. This very short production time is in line with the gluon-bremsstrahlung model of Ref.~\cite{Kop03}.

In this work we discuss different concepts of hadron production and formation in the framework of a probabilistic coupled-channel transport model which allows for a realistic treatment of the FSI. We demonstrate that the coupled-channel effects in the FSI lead to deviations from the results of simple absorption models since particles produced in the FSI affect the low energy part of the hadron spectra and thereby all energy-integrated observables. A strong effect of the coupled channels becomes also visible in the attenuation of kaons since kaon absorption is partially compensated by kaon recreation in the FSI via secondary channels like $\pi +N \rightarrow K+Y+X$. Furthermore, our event-by-event simulation allows us to investigate how the observed attenuation ratios depend on the kinematic cuts and the detector acceptance. We will show that there is a sizable effect of the detector acceptance on the multiplicity ratios, which does not drop out by taking the multiplicity ratio and, therefore, has to be considered explicitly when addressing any robust interpretation of the data.

Being able to compare with almost every observable that is experimentally accessible we can test the limits of a purely hadronic model and figure out at what point an extension becomes necessary. For this reason we also include the EMC data \cite{EMC} on hadron attenuation at 100 and 200 GeV muon energy in our analysis where the hard scale -- set by the photon virtuality $Q^2$ -- is closer to that of the high $p_T$ events at RHIC \cite{RHIC}. We stress that our model is also applicable for a broad range of other reactions such as $\pi A$, $pA$ and $AA$ collisions. A related study of hadron attenuation for high transverse momentum particles in $d+Au$ and $Au+Au$ collisions at RHIC energies has been carried out in \cite{Cass04}.

Our paper is structured in the following way. In Sec.~\ref{sec:gammaN} we discuss the description of the initial (virtual) photon-nucleon interaction using the event generators PYTHIA 6.2 and FRITIOF 7.02 and test our model input with experimental HERMES data on a hydrogen target \cite{Hil03}. Section \ref{sec:shadowing} includes a brief description of the shadowing corrections employed for reactions on nuclei. The BUU transport model -- used for the description of coupled channel FSI -- is recalled in Sec.~\ref{sec:BUU}, while in Sec.~\ref{sec:prehadrons} we discuss our concept of prehadrons. In Sec.~\ref{sec:HERMES} and Sec.~\ref{sec:EMC} we compare the results of our model with experimental data of the HERMES and EMC collaborations. We focus on the multiplicity spectra of charged hadrons as a function of the photon energy $\nu$, the hadron energy fraction $z_h=E_h/\nu$ and the transverse momentum $p_T$. Furthermore, we discuss the effect of the detector geometry and investigate the double-hadron attenuation as well as the attenuation of identified hadrons such as $\pi^\pm$, $\pi^0$, $K^\pm$, $p$ and $\bar{p}$. We close with a short summary in Sec.~\ref{sec:summary}. More technical details of the transport approach are presented in the appendices.

\section{High-energy photon-nucleon interactions}
\label{sec:gammaN} 
Based on the earlier work on high-energy photon-nucleus reactions in \cite{Eff99c} we have developed in Refs.~\cite{Fal03,Eff00,Fal02,incoherent} a method to describe high-energy photon- and electron-induced reactions in the framework of a semi-classical transport model. We thereby split the interaction of the virtual photon with the nucleus into two parts. In the first step i) the virtual photon interacts with a bound nucleon of the target to produce a final state which in step ii) is propagated within the transport model. In the initial $\gamma^*N$ interaction we account for nuclear effects, such as Fermi motion, Pauli blocking and binding energies. Furthermore, coherence length effects in the entrance channel -- leading to nuclear shadowing -- are taken care of using the method developed in Refs.~\cite{Fal02,incoherent}.

For small invariant masses of the photon-nucleon system ($W\leq 2$ GeV) we use an explicit resonance description for the primary (virtual) photon-nucleon interaction as in Ref.~\cite{Leh00}. Above the resonance region $\gamma^*N$ interactions become more complex and in our model the final state is in general determined by the event generator PYTHIA 6.2 \cite{PYTHIA}. We recall that in a high-energy $\gamma^*N$ interaction the virtual photon does not always couple directly to a quark of the target nucleon; depending on its kinematics it may also fluctuate into a vector meson $V=\rho^0,\omega,\phi,J/\psi$ (vector meson dominance (VMD)) or perturbatively branch into a quark-antiquark pair (generalized vector meson dominance (GVMD)) which subsequently scatters off the nucleon. In case of VMD events the same processes as in hadron-hadron interactions can occur including diffractive scattering and hard interactions between the constituents of the nucleon and the vector meson. In addition, most of the hard interactions in PYTHIA involve gluon-bremsstrahlung processes. The high-energy interactions of the photon, therefore, lead to the excitation of several hadronic strings with large invariant masses whose decays are described by the Lund fragmentation \cite{Lund} routine JETSET implemented in PYTHIA.

For technical reasons, resolved $\gamma^*N$ events, i.e.~VMD and GVMD events, can only be generated by PYTHIA if the invariant mass $W$ of the photon-nucleon system is larger than a threshold energy $W_\mathrm{PY}$ which depends on the PYTHIA parameters; it takes the value $W_\mathrm{PY}=4$ GeV for the default parameter set. For invariant masses 2 GeV $\leq W\leq W_\mathrm{PY}$ we employ the event generator FRITIOF 7.02 \cite{FRITIOF} to simulate the resolved photon interactions. As discussed in Appendix~\ref{app:hadron-nucleon} FRITIOF is also used to model the high-energy hadronic FSI in case of nuclear reactions. In contrast to PYTHIA, the FRITIOF model can be applied down to the onset of the resonance region. Below invariant energies of 10 GeV the hadron-hadron interactions in FRITIOF are always diffractive. Throughout this work we employ the same FRITIOF parameters as in Refs.~\cite{Gei98,HSD}. In Ref.~\cite{Gei98} it was shown that this choice of parameters provides a good description of particle production over a broad energy range. The only exception is the suppression factor for $s\bar{s}$ creation in the string fragmentation, which we reset to the FRITIOF default value 0.3 to improve the agreement with more recent strangeness production data \cite{Wag04}.

To simulate the VMD part of the $\gamma^*N$ interaction with FRITIOF we pass the photon as a vector meson $V$ with a probability
\begin{equation}
\label{eq:VMD-prob}
    P_V(W^2,Q^2)=\frac{\left(\frac{W^2}{W^2+Q^2}\right)^3\frac{e^2}{g_V^2}[1+\epsilon \,r(m_V^2,Q^2)]\left(\frac{m_V^2}{m_V^2+Q^2}\right)^2\sigma^\mathrm{tot}_{VN}(W^2)}{\sigma_{\gamma^*N}^\mathrm{tot}(W^2,Q^2)}
\end{equation}
where the numerator, according to Refs.~\cite{PYTHIA,Fri00}, gives the contribution of the vector meson component $V$ to the total photon-nucleon cross section $\sigma_{\gamma^*N}^\mathrm{tot}$. In Eq.~(\ref{eq:VMD-prob}) the VMD coupling constants are denoted as $g_V$; $\sigma_{VN}^\mathrm{tot}$ are the total vector meson nucleon cross sections \cite{PYTHIA,Fri00}
\begin{eqnarray}
\label{eq:vector-cs}
    & &\sigma^\mathrm{tot}_{\rho^0N}(s)\approx\sigma^\mathrm{tot}_{\omega N}(s)\approx 13.63s^\epsilon+31.79s^{-\eta}\,\mathrm{[mb]},\nonumber\\
    & &\sigma^\mathrm{tot}_{\phi N}(s)\approx 10.01s^\epsilon-1.52s^{-\eta}\,\mathrm{[mb]},\nonumber\\
    & &\sigma^\mathrm{tot}_{J/\psi p}(s)\approx\frac{1}{10}\sigma^\mathrm{tot}_{\phi p}(s),
\end{eqnarray}
with $s$ in GeV$^2$ and $\epsilon=0.0808$, $\eta=0.4525$. The factor
\begin{equation}
\label{eq:longitudinal-factor}
    r(m_V^2,Q^2)=0.5\frac{4m_V^2Q^2}{(m_V^2+Q^2)^2}
\end{equation}
accounts for the contribution of longitudinal photons to the photon-nucleon cross section.

\begin{figure}
    \includegraphics[width=13cm]{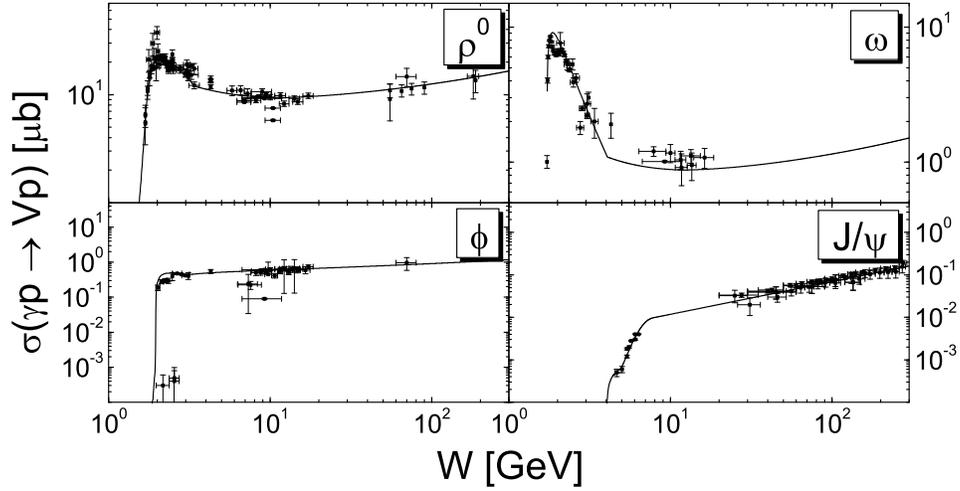}
  \caption{Parameterization of the cross sections for exclusive vector meson photoproduction $\gamma p\to V p$ as a function of the invariant energy $W$ in comparison to the data \cite{PDG98}.}
  \label{fig:v-photoprod}
\end{figure}

In Appendix~\ref{app:hadron-nucleon} we show that FRITIOF insufficiently describes elastic scattering, which directly affects the diffractive vector meson photoproduction $\gamma N\to VN$. In the transition region below $W_\mathrm{PY}$ we, therefore, parameterize the diffractive production cross section for $\rho^0$, $\omega$ and $\phi$ using the Regge prescription of Donnachie and Landshoff \cite{Don00} which also provides the correct dependence on the four-momentum transfer squared $t$. The high-energy cross sections are continuously extrapolated to our parameterization in the resonance region (cf.~Ref.~\cite{Eff99c}). For the description of $J/\psi$ production close to threshold we use the two and three gluon exchange model of Brodsky et al.~\cite{Bro00} and parameterize the high-energy $\sqrt{s}$ dependence as
\begin{equation}
\label{eq:jpsi}
    \sigma_{\gamma N\to J/\psi N}=0.002 \left(\frac{\sqrt{s}}{\mathrm{GeV}}\right)^{0.77}\,\mu \mathrm{b}.
\end{equation}
Fig.~\ref{fig:v-photoprod} shows our parameterization of diffractive vector meson production for $\rho^0$, $\omega$, $\phi$ and $J/\psi$ in comparison to a collection of experimental data and demonstrates that our 'input' for the $\gamma^*p$ reaction is well in accordance with experimental information.

For the processes $\gamma N\to V\Delta$ we use our parameterizations of the differential cross section for $\gamma N\to \phi \Delta$ and $\gamma N\to \omega \Delta$ from Ref.~\cite{Muehlich} and take the same matrix element for $\gamma N\to \rho^0\Delta$ as for $\gamma N\to \omega \Delta$.

To account for the virtuality $Q^2$ of the photon we multiply the vector meson photoproduction cross sections with the VMD formfactor of Ref.~\cite{Fri00}, which is also used in the PYTHIA model (cf.~Eq.~(\ref{eq:VMD-prob}))
\begin{equation}
\label{eq:electro-formfactor}
    F_V^2(Q^2)=\left(\frac{W^2}{Q^2+W^2}\right)^3[1+\epsilon \,r(m_V^2,Q^2)]\left(\frac{m_V^2}{m_V^2+Q^2}\right)^2.
\end{equation}

The FRITIOF model has the tendency to simply flip the spin of the incoming vector meson to zero thereby creating a lot of events like $\gamma^*N\to \pi^0N$ with vanishing momentum transfer, a problem not present in PYTHIA. These events would correspond to diffractive $\pi^0$ production. However, since the Pomeron carries the quantum numbers of the vacuum it cannot change the charge conjugation quantum number of the incoming $\rho^0$ ($C=-1$) to that of the $\pi^0$ ($C=+1$). We, therefore, simply remove these unphysical events from our simulation.

We have compared the particle spectra of VMD events as generated by FRITIOF and PYTHIA at the threshold energy $W_\mathrm{PY}=4$ GeV and found no significant differences between the two approaches. However, we keep the option to also generate the VMD events above $W_\mathrm{PY}$ = 4 GeV by FRITIOF to ensure a continuous transition at invariant energies $W>W_\mathrm{PY}$. For calculations at EMC energies we always use the PYTHIA model for event generation. Note, however, that the VMD contribution vanishes with increasing $Q^2$ and already at $Q^2=1$ GeV$^2$ the VMD part $\sigma_\mathrm{VMD}^{\gamma^*N}$ accounts for only 50\% of the total photon-nucleon cross section at HERMES energies \cite{incoherent}.

The GVMD part below $W_\mathrm{PY}$ is modeled by passing a vector meson with the quark content of the $q\bar{q}$ fluctuation to FRITIOF. In analogy to Eq.~(\ref{eq:VMD-prob}) the probability is now given by the corresponding GVMD term of Ref.~\cite{Fri00} normalized to $\sigma^\mathrm{tot}_{\gamma^*N}$. As discussed in Ref.~\cite{Fri00} the products of a GVMD and VMD event are different in nature since the VMD and GVMD components of the photon involve different intrinsic transverse momenta. However, we have shown in Ref.~\cite{incoherent} that the GVMD part in the transition region only contributes by less than 15\% to the total photon-nucleon cross section. Hence, one can postpone a more sophisticated treatment unless one concentrates on events that are especially triggered by the GVMD component of the photon.

\begin{figure}
    \includegraphics[width=11cm]{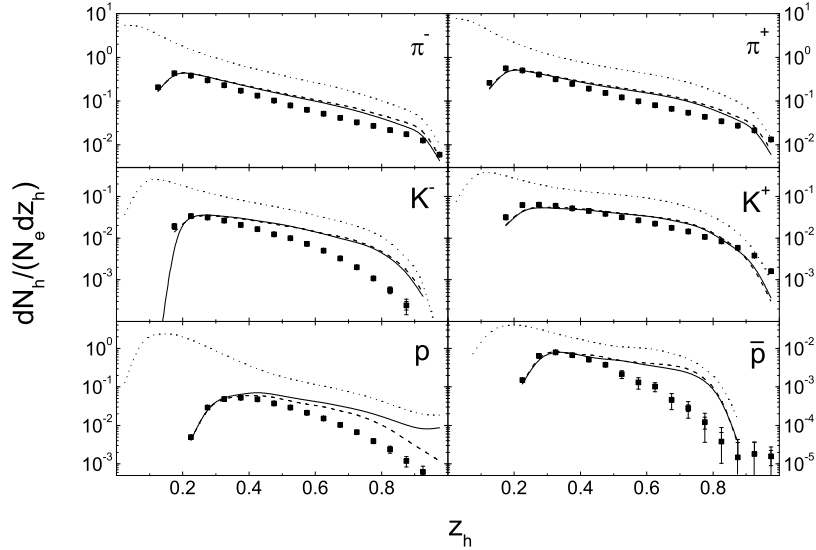}
  \caption{Energy spectra of $\pi^\pm$, $K^\pm$, $p$ and $\bar{p}$ within the HERMES detector acceptance for a hydrogen target. The spectra are normalized to the number of deep inelastically scattered leptons. The solid lines represent the calculated results using our standard method of generating all resolved photon-nucleon events below $W_\mathrm{PY}=4$ GeV and all VMD events above $W_\mathrm{PY}$ with FRITIOF. Generating all events down to $W=3$ GeV with PYTHIA leads to the spectra shown in terms of the dashed lines. The dotted lines show the result of a simulation without employing any kinematic or detector cuts on the hadrons. The data (full squares) are taken from Ref.~\cite{Hil03}.}
  \label{fig:elementary}
\end{figure}

We now compare our results for electroproduction on a proton target with the hydrogen data of the HERMES collaboration \cite{Hil03} taken at a positron beam energy $E_\mathrm{beam}=27.6$ GeV. In Fig.~\ref{fig:elementary} we show the $z_h = E_h/\nu$ spectra $dN_h/dz_h$ of $\pi^\pm$, $K^\pm$, $p$ and $\bar{p}$ normalized to the number $N_e$ of deep inelastically scattered positrons. In our calculation we apply the same cuts on the event kinematics as in the experiment: For the positrons these are $Q^2>1$ GeV$^2$, $W^2>10$ GeV$^2$ and $0.1<y=\nu/E_\mathrm{beam}<0.85$. For the hadrons we require 1 GeV/$c$ $<p_\pi<15$ GeV/$c$, 2 GeV/$c$ $<p_{K,p,\bar{p}}<15$ GeV/$c$ as well as $x_F>0.1$, where we define the Feynman variable $x_F$ as in experiment by
\begin{equation}
    x_F=\frac{p_\parallel^\mathrm{cm}}{|\vec q_\mathrm{cm}\,|}.
\end{equation}
Here $p_\parallel^\mathrm{cm}$ denotes the momentum of the hadron parallel to the momentum $\vec q_\mathrm{cm}$ of the virtual photon in the center-of-mass frame of the photon-nucleon system. Since the data are not acceptance and efficiency corrected we account for the angular acceptance of the HERMES detector \cite{acceptance}, i.e.~$\pm 170$ mrad horizontally and $\pm(40-140)$ mrad vertically, for both the scattered positrons and the produced hadrons in our simulation. The solid line in Fig.~\ref{fig:elementary} shows our result using FRITIOF for the resolved events below $W_\mathrm{PY}=4$ GeV as well as for all VMD events above $W_\mathrm{PY}$. The dashed line represents a calculation where we have changed the default parameters of PYTHIA 6.2 in such a way that it is applicable down to $W_\mathrm{PY}\approx3$ GeV and, hence, simulate {\it all} events with PYTHIA. Except for a small deviation in the proton spectra both methods yield essentially the same result. In view of the fact, that we do not include the detector efficiency in our calculation, which is unknown to the authors, but only account for its angular acceptance, our calculations are in satisfying agreement with the experimental data. One can see that without any further fine tuning our simulation including the kinematic cuts and detector acceptance reproduces the absolute size of the multiplicity spectra. We have also compared the $x_F$, $p_T$ and total momentum-spectra of the different hadron species with the HERMES data and find a similar good agreement. To demonstrate the effects of the kinematic cuts and the limited angular acceptance of the HERMES detector, the dotted line in Fig.~\ref{fig:elementary} shows the results of a simulation where no cuts on the hadron kinematics have been applied and where we have assumed a $4\pi$-detector.

\section{Nuclear Shadowing}
\label{sec:shadowing}
As discussed in the previous section the photon does not always interact directly with a quark inside the target nucleon since its wavefunction is a superposition of a bare photon and resolved hadronic fluctuations:
\begin{equation}
\label{eq:had}
|\gamma\rangle=c_\mathrm{bare}|\gamma_\mathrm{bare}\rangle+
\sum_{V=\rho^0,\omega,\phi,J/\psi}c_V|V\rangle+\sum_{q=u,d,s,c,b}c_q|q\bar{q}\rangle.
\end{equation}
In an electron-nucleus interaction the hadronic components of the photon in (\ref{eq:had}) are shadowed on their way through the nucleus due to the strong interaction of these components with the hadronic environment. The strength of this shadowing effect depends on the 'coherence length', i.e.~the distance $l_h$ that the photon travels as a hadronic fluctuation $h$. This distance can be estimated via the uncertainty principle:
\begin{equation}
\label{eq:coherence-length}
    l_h=q_h^{-1}=(k-k_h)^{-1} ,
\end{equation}
where $k=\sqrt{\nu^2+Q^2}$ is the photon momentum and $k_h=\sqrt{\nu^2-m_h^2}$ denotes the momentum of the hadronic fluctuation  on its mass shell. If the coherence length $l_h$ becomes larger than the mean-free-path of the hadronic fluctuation in the nuclear medium
the hadronic fluctuation gets shadowed. In the kinematic region of interest in this work the shadowing of the massive $q\bar{q}$ fluctuations can be neglected and one is only left with the shadowing of the photon's VMD components. In Ref.~\cite{Fal02,incoherent} we have calculated the modification of the vector meson components at position $\vec r=(\vec b,z)$ in the nucleus:
\begin{equation*}
|V\rangle\to\left(1-\overline{\Gamma^{(A)}_V}(\vec
r)\right)|V\rangle,
\end{equation*}
where the nuclear profile function $\Gamma^{(A)}_V$ is determined within Glauber theory, i.e.
\begin{eqnarray}
  \overline{\Gamma^{(A)}_V}(\vec b,z)&=&\intop_{-\infty}^{z}dz_i\rho(\vec b,z_i)\frac{\sigma_{VN}^\mathrm{tot}}{2}(1-i\alpha_V)e^{iq_V(z_i-z)}\nonumber\\
  & &\quad\times\exp\left[-\frac{1}{2}\sigma_{VN}^\mathrm{tot}(1-i\alpha_V)\intop_{z_i}^{z}dz_k\rho(\vec b,z_k)\right]. \label{eq:av_profile}
\end{eqnarray}
Here $\sigma_{VN}^\mathrm{tot}$ denotes the total vector meson-nucleon cross section and $\alpha_V$ is the ratio of the real and imaginary part of the $VN$ forward scattering amplitude. Note that the coherence length (\ref{eq:coherence-length}) enters in the phase factor of Eq.~(\ref{eq:av_profile}). The modification of the photon's vector meson components are then taken into account
when generating the scattering event with PYTHIA and FRITIOF. As we have demonstrated in Ref.~\cite{incoherent} our prescription is in full agreement with the coherence length effects observed in incoherent $\rho^0$ production off nuclei.
\section{BUU transport model}
\label{sec:BUU} 
Our transport model -- employed for the description of FSI -- is based on the Boltzmann-Uehling-Uhlenbeck (BUU) equation. It has been originally developed to model heavy-ion collisions at SIS energies \cite{HI} and later been extended to describe pion-induced reactions \cite{pion} as well as photo- and electroproduction \cite{Leh00} in the resonance region. In the more recent past we have extended the model to simulate also nuclear interactions of high-energy photons and electrons in the kinematic regime of the Jefferson Lab \cite{Eff99c,Eff00,Fal02} and HERMES \cite{incoherent,Fal03} experiments. Obviously, one of the advantages of the model is its applicability to many different nuclear reactions with the same set of parameters and physics assumptions. In addition, the coupled-channel treatment of the FSI goes far beyond the standard Glauber approach since it allows for a side-feeding of channels under study and not just absorption.

The BUU transport model is based on a set of generalized transport equations for each particle species $i$,
\begin{equation}
\label{eq:BUU}
    \left(\frac{\partial}{\partial t}+\vec \nabla_{\vec p} H\vec \nabla_{\vec r}
    -\vec \nabla_{\vec r} H\vec \nabla_{\vec p}\right) F_i(\vec r,\vec p,\mu;t)=
    I_\mathrm{coll}(\{F_j\})
\end{equation}
where
\begin{equation}
\label{eq:hamilton}
    H=\sqrt{(\mu+U_S(\vec r,\vec p;t))^2+{\vec p}^2}
\end{equation}
denotes the relativistic Hamilton function of a particle with mass $\mu$ in a scalar potential $U_S$. For vanishing collision term $I_\mathrm{coll}$ Eq.~(\ref{eq:BUU}) describes the time evolution of the spectral phase-space density
\begin{equation}
\label{eq:spectral-phase-space}
    F_i(\vec r,\vec p,\mu;t)=f_i(\vec r,\vec p;t){\mathcal A_i(\mu,\vec p)}
\end{equation}
of non-interacting particles that move in a scalar mean-field potential $U_S$, which in general depends on the phase-space densities $\{f_j\}$ of all other particle species (including $i$). The spectral function ${\mathcal A}_i$ of particle type $i$ is parameterized as in Ref.~\cite{Eff99c}:
\begin{equation}
\label{eq:spectral-function}
    {\mathcal A}_i(\mu)=\frac{2}{\pi}\frac{\mu^2 \Gamma_\mathrm{tot}(\mu,\vec p)}{(\mu^2-M_i^2)^2+\mu^2\Gamma_\mathrm{tot}^2(\mu,\vec p)},
\end{equation}
where $M_i$ denotes the pole mass of particle $i$ and $\Gamma_\mathrm{tot}$ its total width which in medium also depends on the particle momentum $\vec p$.

The collision term
\begin{equation}
\label{eq:coll-term}
    I_\mathrm{coll}(\{F_j\})=-i\Sigma^>_i(\{F_j\}){\mathcal A_i}f_i-i\Sigma^<_i(\{F_j\}){\mathcal A_i}(1\pm f_i)
\end{equation}
in Eq.~(\ref{eq:BUU}) accounts for changes in the spectral phase-space density of particle species $i=a_1$ due to multi-particle collisions of the type $a_1,\ldots,a_n\to b_1\ldots b_m$ and $b_1\ldots b_m\to a_1,\ldots,a_n$. The factor $(1\pm f)$ has the plus sign for bosons (Bose enhancement) and the minus sign for fermions (Pauli blocking). The collision term consists of a loss term for particle species $a_1$
\begin{eqnarray}  
\label{loss}
i\Sigma^>_{a_1}&=&\frac{1}{2E_{a_1}}\intop\left(\prod_{j=2}^n
g_{a_j}\frac{d^3p_{a_j}}{(2\pi)^3}\frac{d\mu_{a_j}}{2E_{a_j}}{\mathcal
A}_{a_j} (\mu_{a_j},\vec p_{a_j})f_{a_j}\right)\\ & &
\quad\times\left(\prod_{k=1}^m\frac{d^3p_{b_k}}{(2\pi)^3}\frac{d\mu_{b_k}}{2E_{b_k}}{\mathcal
A}_{b_k}(\mu_{b_k},\vec p_{b_k})(1\pm f_{b_k})\right)\nonumber\\
    & &\qquad\times(2\pi)^4\delta^4(\sum_{j=1}^np_{a_j}
    -\sum_{k=1}^mp_{b_k})S_{a_2,\ldots, a_n}S_{b_1,
    \ldots, b_m}\overline{|{\mathcal M}_{a_1,\ldots,a_n\to b_1\ldots b_m}|^2}\nonumber
\end{eqnarray}
and a gain term
\begin{eqnarray} \label{gain}
-i\Sigma^<_{a_1}&=&\frac{1}{2E_{a_1}}\intop\left(\prod_{j=1}^m
g_{b_j}
\frac{d^3p_{b_j}}{(2\pi)^3}\frac{d\mu_{b_j}}{2E_{b_j}}{\mathcal
A}_{b_j} (\mu_{b_j},\vec p_{b_j})f_{b_j}\right) \\ & &
\quad\times\left(\prod_{k=2}^n\frac{d^3p_{a_k}}{(2\pi)^3}\frac{d\mu_{a_k}}{2E_{a_k}}{\mathcal
A}_{a_k}(\mu_{a_k},\vec p_{a_k})(1\pm f_{a_k})\right)\nonumber\\ &
& \qquad\times(2\pi)^4\delta^4(\sum_{j=1}^mp_{b_j}-\sum_{k=1}^n
p_{a_j})S_{b_1,\ldots, b_m}S_{a_2,\ldots, b_n}\overline{|{\mathcal
M}_{b_1,\ldots,b_m\to a_1\ldots a_n}|^2}\nonumber
\end{eqnarray}
where $g_j$ are degeneration factors that account for the spin of particle $j$. The quantity $\overline{|{\mathcal M}|^2}$ is the squared matrix element averaged over incoming and summed over outgoing spins. $S$ denotes symmetry factors that take into account the number of identical particles in the incoming and outgoing channel, e.g.,
\begin{equation}
    S_{b_1,\ldots, b_m}=\prod_{k=1}^m\frac{1}{M_{b_k}!}
\end{equation}
with $M_{b_k}$ denoting the multiplicity of particle $b_k$.

All transport equations (\ref{eq:BUU}) are coupled by the collision term and the scalar potential $U_S$ in the Hamilton function (\ref{eq:hamilton}) where in our model the latter is only incorporated in case of baryons. Furthermore, in the general collision term we will restrict to $2 \rightarrow n$ transitions ($n\geq 1$) which is sufficient for the lepton-nucleus reactions of interest. For a specification of the scalar potential $U_S$ in case of baryons as well the $2 \rightarrow n$ transition rates we refer the reader to Appendix~\ref{app:BUU-details} of this work.

\section{Prehadrons}
\label{sec:prehadrons}
In this section we compare the concept of 'prehadrons' in the Lund model \cite{Lund} with that generally used in transport models \cite{Eff00,HSD,UrQMD}. The two concepts mainly differ in the treatment of the production proper time $\tau_p$ of the color neutral prehadrons which are created in the string decay. While the production time of 'prehadrons' in the Lund model depends on the energy and momentum of the fragments, $\tau_p$ is  set to zero in conventional transport approaches \cite{Eff00,HSD,UrQMD}. In order to specify the differences and to define the relevant quantities (production and formation times) we give a brief reminder of the Lund model in the following.

\subsection{Prehadrons in the Lund model}
\label{ssec:Lund-prehadrons}
\begin{figure}
    \includegraphics[width=12cm]{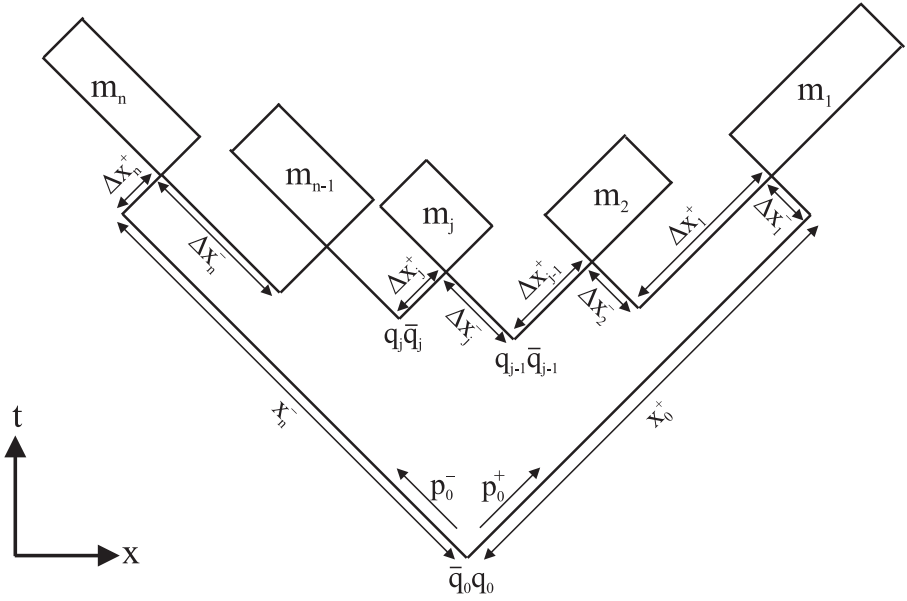}
  \caption{Simplified space-time picture of the fragmentation of a $\bar{q}_0q_0$ string with invariant mass squared $W^2=p_0^+p_0^-$. The string decays into $n$ hadrons with masses $m_1,\ldots,m_n$ due to the creation of $n-1$ quark-antiquark pairs from the vacuum. The production proper times $\tau_p(j-1)$ and $\tau_p(j)$ correspond to the vertices where the $q_{j-1}\bar{q}_{j-1}$ and the $q_j\bar{q}_j$ pairs are produced. The formation proper time $\tau_f(j)$ of hadron $j$ refers to the space-time point where the world lines of the quark $q_{j-1}$ and the antiquark $\bar{q}_j$ cross for the first time. Note that, the 'production times' of the $q_0$ and $\bar{q}_0$ -- which in the end are contained in the first and the last rank hadrons -- are always zero.}
   \label{fig:lund}
\end{figure}

According to the Lund model \cite{Lund} the hadronic strings -- created in the high-energy interactions -- decay due to the creation of quark-antiquark pairs from the vacuum. Fig.~\ref{fig:lund} shows the space-time picture of a Lund fragmentation process in the rest frame of a $\bar{q}_0q_0$ string. In our example we neglect the masses of the quarks, i.e.~all quarks move on the lightcone. At time zero the quark $q_0$ carries the lightcone momentum $p_0^+$ while the antiquark $\bar{q}_0$ carries the lightcone momentum $p_0^-$. During their separation they loose energy and momentum to the string until they reach their turning points:
\begin{equation}
\label{eq:turning1}
    x_0^+=\frac{p_0^+}{\kappa}\, ,\qquad\qquad x_0^-=0
\end{equation}
and
\begin{equation}
\label{eq:turning2}
    x_n^+=0\, , \qquad\qquad x_n^-=\frac{p_0^-}{\kappa},
\end{equation}
where the string tension $\kappa$ in vacuum is known to be $\kappa\approx 1$ GeV/fm.

The production process of the $n$ mesons with masses $m_1,\ldots,m_n$ due to the creation of $n$-1 quark-antiquark pairs $q_j\bar{q}_j$ at the production vertices $x_j^\pm$ can be viewed as a series of steps along the (positive) lightcone. Starting at the turning point (\ref{eq:turning1}) of the original quark $q_0$ one takes random steps along the positive lightcone
\begin{equation}
\label{eq:random-step}
    \Delta x_j^+=z_jx_{j-1}^+\, , \qquad z_j\in[0,1]
\end{equation}
where the probability distribution for the random variable $z_j$ is given by the Lund fragmentation function
\begin{equation}
\label{eq:Lund-fragmentation}
    f(z_j)=N\frac{(1-z_j)^a}{z_j}\exp(-b\frac{m_{Tj}^2}{z_j}) ,
\end{equation}
which depends on the transverse mass
\begin{equation}
    m_{Tj}=\sqrt{m_j^2+p_T^2}
\end{equation}
of the produced hadron. A step along the negative lightcone is fixed by the on-shell condition for hadron $j$
\begin{equation}
\label{eq:onshell-step}
        \Delta x_j^-=-\frac{m_{Tj}^2}{\kappa^2\Delta x_j^+}.
\end{equation}
Thus we obtain the recursion formulae
\begin{eqnarray}
x_j^+&=&x_{j-1}^+-\Delta x_j^+=(1-z_j)x_{j-1}^+\label{eq:lund-recursion1}\\
x_j^-&=&x_{j-1}^--\Delta x_j^-=x_{j-1}^-+\frac{1-z_j}{z_jx_j^+}\frac{m_{Tj}^2}{\kappa^2}\label{eq:lund-recursion2}
\end{eqnarray}
together with the constraint that the path must end at the turning point (\ref{eq:turning2}) of $\bar{q}_0$.

Obviously, the proper times $\tau_p(j)$ of the $q_j\bar{q}_j$ production vertices
\begin{equation}
    \tau_p^2(j)=x_j^+x_j^-
\end{equation}
strongly depend on the mass $m_j$ of the produced hadron. The same holds true for the proper time of crossing world lines for the pair $q_{j-1}$ and $\bar{q}_j$
\begin{equation}
    \tau_f^2(j)=x_{j-1}^+x_j^-.
\end{equation}
The latter is often identified with the formation time $\tau_f(j)$ of the hadron $j$.

In the Lund picture there are in principle three timescales involved in the production of the hadron $(\bar{q}_jq_{j-1})$. This is the formation time $\tau_f(j)$ of the hadron as well as the times $\tau_p(j-1)$ and $\tau_p(j)$ when the first and the second constituent
parton is produced. As soon as the $q_{j-1}$ and the $\bar{q}_j$ are created they form an intermediate object which can be viewed as a color neutral prehadron. As pointed out in Refs.~\cite{Gei99,Cio02} the longitudinal dimension of the string and its (color neutral)
fragments is never larger than typically 1 fm. Therefore, it does not make much sense to talk about freely propagating colored quarks between $\tau_p(j-1)$ and $\tau_p(j)$ since each color charge is always attached via a string to the nearby anticharge. Instead one has to deal with the propagation of color neutral prehadrons and color neutral remainder strings during the fragmentation process.

As an example we recall the production and formation times of the first rank particle (containing the $q_0$),
\begin{eqnarray}
\label{eq:production-time}
    \tau_p^2&=&\frac{m_T^2}{\kappa^2}\frac{1-z}{z} , \\
\label{eq:formation-time}
    \tau_f^2&=&\frac{m_T^2}{\kappa^2}\frac{1}{z} ,
\end{eqnarray}
where for the first rank particle $z$ corresponds to the fraction $z_h^+$ of the positive lightcone momentum $p_0^+$ that is carried away by the first rank hadron. For fixed $z$ both times increase linearly with the (transverse) hadron mass $m_T$ which implies that heavier hadrons are in general created later in time. Note, however, that the mass also enters into the Lund fragmentation function (\ref{eq:Lund-fragmentation}). Furthermore, the average production and formation proper times increase with the rank of the hadron as long as one considers a fragmentation into infinitely many hadrons. We stress that the assumption of infinitely many fragments enters explicitly in the derivation of the production and formation times in Ref.~\cite{Bia87} which are averaged over infinitely many ranks.
At HERMES energies, however, a string typically fragments into only 3--5 hadrons. Thus, when using the Lund production and formation times in our simulation, we explicitly extract the space-time points of the vertices and crossing points of pairs from the
JETSET routines for each fragmenting string since the limit addressed in  Ref.~\cite{Bia87} is inappropriate for the kinematical regime of interest.

\begin{figure}
    \includegraphics[width=15cm]{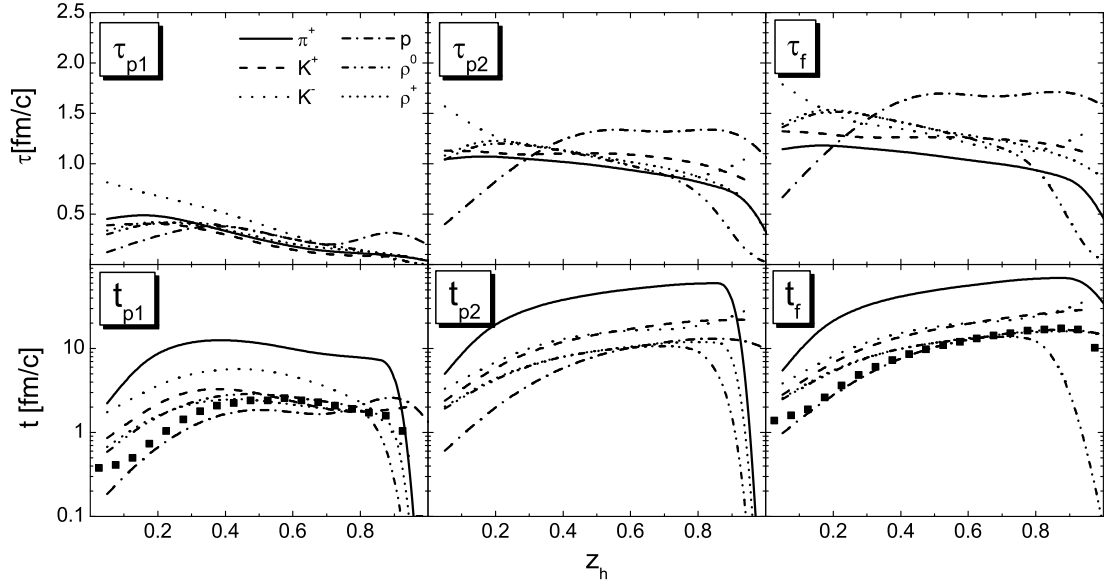}
  \caption{Average (proper) times for hadron formation and the production of the first and second constituent as a function of $z_h=E_h/\nu$ in the kinematic regime of the HERMES experiment. All proper times have been extracted directly from the JETSET routines which describe the string fragmentation in PYTHIA. The solid squares indicate the average starting times of the prehadronic and hadronic interactions in the constituent quark concept (\ref{eq:prehadrons}) using a formation time $\tau_f=0.5$ fm/$c$.}
\label{fig:JETSET2}
\end{figure}

In the upper panel of Fig.~\ref{fig:JETSET2} we show the $z_h$ dependence of the extracted average production proper times of the two hadron constituents as well as the hadron formation proper time $\tau_f$ in electron-nucleon scattering at HERMES energies. We always
label the smaller production time $\tau_{p1}$ and the large $\tau_{p2}$. Note that, if more than one string is produced in the electron-nucleon interaction, the observed $z_h=E_h/\nu$ is in general different from $z_h^+$. The difference is most significant for
$z_h\lesssim0.5$. However, one also has to be careful with the interpretation of $z_h$ in the limit $z_h\to 1$. In this kinematic regime the hadron spectrum is dominated by diffractively produced vector mesons (cf.~Fig.~\ref{fig:leading}) whose production times
$\tau_{p1}$ and $\tau_{p2}$ for the constituents are both zero. Furthermore, we have learned from our theoretical study of incoherent $\rho^0$ electroproduction \cite{incoherent} that also the formation time $\tau_f$ of a diffractively produced
vector meson is compatible with zero in the HERMES kinematical regime. Consequently, the proper times $\tau_{p1}$, $\tau_{p2}$ and $\tau_f$ for the $\rho^0$ (dash-dot-dotted line) vanish as $z_{\rho^0}\to 1$. 

In a direct photon-nucleon interaction the probability, that a meson contains the struck quark from the string end, increases with $z_h$ and, consequently, $\tau_{p1}$ vanishes for larger $z_h$. On the other hand, the diquark that is left behind, i.e.~the second string end, will most likely form a proton with small $z_h$ which leads to a vanishing of $\tau_{p1}$ for low-$z_h$ protons (dash-dotted line). In case of diffractive photon-nucleon interactions the proton stays intact and $\tau_{p1}$, $\tau_{p2}$ and $\tau_f$ are zero which leads to a decrease of $\tau_{p2}$ and $\tau_f$ for low-$z_h$ protons. The Lund model predicts a vanishing of the production time $\tau_{p2}$ for $z\to 1$ which can be seen from Eq.~(\ref{eq:production-time}). However, because of the finite bin-size the production time $\tau_{p2}$ in Fig.~\ref{fig:JETSET2} stays finite in the last $z_h$ bin for all hadrons except for the diffractively produced vector mesons.

In fact, one observes the general tendency that  more massive hadrons like protons and $\rho$ mesons are in general formed later in proper time than lighter hadrons like pions (solid line). This was already indicated by Eqs.~(\ref{eq:production-time}) and (\ref{eq:formation-time}) for the production and formation times of the first rank hadron in a string fragmentation. Again the diffractively produced vector mesons are the exception.

We find that the production proper time $\tau_{p1}$ of the first hadron constituent -- averaged over all hadrons -- lies between 0.04 fm/$c$ at $z_h\approx 0.95$ and 0.4 fm/$c$ at $z_h\approx0.35$. The production proper time $\tau_{p2}$ of the second constituent is somewhat larger. It ranges from 0.3 fm/$c$ at $z_h\approx 0.95$ to about 1.2 fm/$c$ at $z_h\approx0.35$. The average hadron formation proper time $\tau_f$ is of the order 1.1 -- 1.5 fm/$c$ except for the very large-$z_h$ hadrons which are dominated by diffractively produced $\rho^0$ mesons and, hence, have again a very small formation time around 0.4 fm/$c$ at $z_h\approx 0.95$. 

In the lower panel of Fig.~\ref{fig:JETSET2} we also show the corresponding production and formation times in the laboratory frame, i.e.~the rest frame of the target nucleon:
\begin{eqnarray}
 t_p&=&\gamma\cdot\tau_p=\frac{z_h\nu}{m_h}\cdot\tau_p\label{eq:lab-production-time}\\
 t_f&=&\gamma\cdot\tau_f=\frac{z_h\nu}{m_h}\cdot\tau_f\label{eq:lab-formation-time}.
\end{eqnarray}
We emphasize that simply boosting the proper times with the Lorentz gamma of the hadron -- as done in Eqs.~(\ref{eq:lab-production-time}) and (\ref{eq:lab-formation-time}) -- is just an approximation. We currently carry out a more thorough investigation on this topic \cite{Gal04}. Because of time dilatation the light pions now have average production and formation times as large as 10--70 fm/$c$. However, as we discuss in the next section (and as can be seen from Fig.~\ref{fig:leading}) pion production receives a large contribution from string fragmentation into $\rho$ mesons that -- due to their larger mass -- have a considerably smaller average production time 0.5--3 fm/$c$ in the laboratory frame.

Note that the production and formation times in Fig.~\ref{fig:JETSET2} are {\it averaged} times. In our actual simulation we assign in each scattering event the {\it individual} production and formation time from JETSET for each hadron.

\subsection{Prehadrons in transport models}
\label{ssec:transport-prehadrons}
The hadrons containing the endpoint partons $q_0$ and $\bar{q}_0$ of the string, that emerges from the excitation of a hadron in a scattering event, are not entirely created from the vacuum as the other $q\bar{q}$ pairs. Indeed, the hadrons that emerge from the string ends can be viewed as the remnants of the original hadron. In the transport models \cite{Eff00,HSD,UrQMD} each string decay into color neutral prehadrons is determined instantaneously at the production vertex. During the formation time $\tau_f$ only the beam and target remnants are allowed to interact with the nuclear environment whereas the hadrons containing only quarks or antiquarks from the vacuum are propagated freely without interactions up to their hadron formation time in the computational reference frame.

It is hard to identify the beam and target remnants of a $\gamma^*N$ event because particle production in PYTHIA and FRITIOF is complicated by the resolved photon interactions as well as the initial and final state gluon radiation that occur in many hard scattering events. As a result one often ends up with more than a single string per DIS event or a string that contains one or several radiated gluons, i.e.~it might have the structure $qg\bar{q}$, $qgg\bar{q}$ etc. In general, the gluons have momentum components transverse to the cm momentum of the quark and antiquark. This leads to more complicated string topologies than just a linearly expanding string \cite{Lund} and has to be taken into account in the fragmentation.

To identify the target and beam remnants of a binary collision we, therefore, trace the (anti-)quarks of the projectile and target all the way through the fragmentation process in JETSET. At the end of the reaction we then know those hadrons that contain the original (anti-)quarks, i.e.~the beam and target remnants. Correspondingly, also those hadrons exclusively made from (anti-)quarks created in the string fragmentation are known explicitly. 

In Fig.~\ref{fig:leading} we show the energy spectra of hadrons produced in electron-nucleon interactions at HERMES. The different lines indicate the hadrons that contain zero (solid line), one (dashed line), two (dotted line) or three quarks (dash-dotted line) from the beam or target. In the dotted curves one can clearly identify the diffractive peak in the $\rho^0$ and $\phi$ spectra at $z_h\approx 1$, where the vector mesons contain the $q\bar{q}$ of the resolved photon. By comparing the $\rho^0$ and $\pi^0$ spectra one finds that the $\rho^0$ strongly contributes to the total pion yield by its subsequent decay into $\pi^+\pi^-$. From the $K^+$ and $K^-$ spectra one sees (dashed lines) that due to their quark content ($\bar{u}s$) the $K^-$ mesons contain less quarks from the beam or target than the $K^+$ mesons. The $K^-$, that are not solely made of quarks and antiquarks created from the vacuum in the string fragmentation, carry (anti-)quarks from the resolved photon component or the nucleon sea. Finally, one finds that there are only very few protons containing no quarks from the beam or target since diquark-antidiquark creation is strongly suppressed in the string fragmentation due to the relatively large diquark masses. This also explains why most protons at large $z_h$ contain two of the original quarks, i.e.~the diquark from the target nucleon struck by the photon.

In our default approach we set the production times $\tau_p$ of all prehadrons to zero and rescale their cross sections during the formation time $\tau_f$ according to the constituent quark model:
\begin{eqnarray}
\label{eq:prehadrons}
\sigma_\mathrm{prebaryon}^\mathrm{tot}&=
&\frac{n_\mathrm{org}}{3}\sigma_\mathrm{baryon} , \nonumber\\
    \sigma_\mathrm{premeson}^\mathrm{tot}&=&\frac{n_\mathrm{org}}{2}\sigma_\mathrm{meson}
    ,
\end{eqnarray}
where $n_\mathrm{org}$ denotes the number of (anti-)quarks in the prehadron stemming from the beam or target. As a consequence the prehadrons that solely contain (anti-)quarks produced from the vacuum in the string fragmentation do not interact during $\tau_f$. The
assumption (\ref{eq:prehadrons}) is guided by the constraints of unitarity, which implies that the summed cross section of the products from a scattering process should not exceed the cross section for the initial hadrons on short time scales. For simplicity we assume
that the formation time is a constant $\tau_f$ in the rest frame of each hadron and that it does not depend on the particle species as in Refs.~\cite{Gei98,HSD}. In Section \ref{sec:EMC} we also discuss other concepts of the prehadronic cross section. However, as we will show in Sec.~\ref{sec:HERMES} the 'standard' concept (\ref{eq:prehadrons}) already gives a good description of the available HERMES data. This also holds for particle production in proton-nucleus and nucleus-nucleus collisions in a wide dynamical range \cite{brat}.

\section{Hadron attenuation at HERMES energies}
\label{sec:HERMES}
\begin{figure}
    \includegraphics[width=14cm]{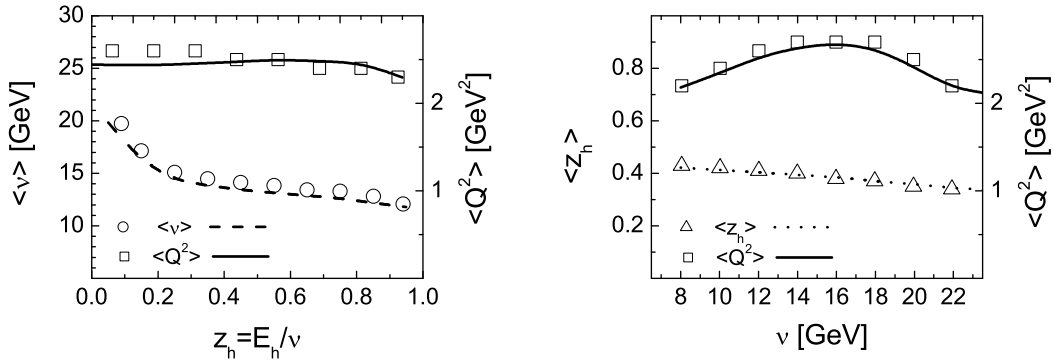}
  \caption{Model predictions for the average values of the kinematic variables in charged hadron production off $^{84}$Kr in comparison with the experimental numbers at HERMES \cite{Muc03}. For the calculation we used the formation time $\tau_f=0.5$ fm/$c$ and the constituent quark concept (\ref{eq:prehadrons}) for the prehadronic cross sections. {\it Left:} $\langle\nu\rangle$ and $\langle Q^2\rangle$ as a function of $z_h$ compared to the experimental values for a $^{84}$Kr target (open symbols). {\it Right:} Same for $\langle z_h\rangle$ and $\langle Q^2\rangle$ as a function of $\nu$.} \label{fig:average}
\end{figure}

We start with an investigation of charged hadron attenuation in semi-inclusive DIS of 27.6 GeV positrons off nitrogen and krypton. We apply the same kinematic cuts as in experiment \cite{HERMESDIS_new}: $Q^2>1$ GeV$^2$, $W>2$ GeV, $\nu>7$ GeV, $y<0.85$, $x>0.02$ and $E_h>1.4$ GeV as well as the cut $z_h>0.2$ for the $\nu$ and $p_T$ spectra. Furthermore, we account for the angular acceptance of the HERMES detector. The left panel in Fig.~\ref{fig:average} shows the average values of the photon energy $\nu$ and the virtuality $Q^2$ as a function of the energy fraction $z_h$ of the charged hadrons produced on a $^{84}$Kr target; the right panel shows the average values of $z_h$ and $Q^2$ as a function of $\nu$. In the simulation we used the prehadron concept (\ref{eq:prehadrons}) and a formation time $\tau_f=0.5$ fm/$c$. Obviously, our simulation is in perfect agreement with the experimental values for the average kinematic variables (open symbols).

The observable of interest is the multiplicity ratio defined as:
\begin{equation}
\label{eq:multiplicity-ratio}
R_M^h(z_h,\nu,p_T^2,Q^2)=\frac{\frac{N_h(z_h,\nu,p_T^2,Q^2)}
{N_e(\nu,Q^2)}\big|_A}{\frac{N_h(z_h,\nu,p_T^2,Q^2)}{N_e(\nu,Q^2)}\big|_D}
,
\end{equation}
where $N_h$ is the yield of semi-inclusive hadrons in a given $(z_h,\nu,p_T^2,Q^2)$-bin and $N_e$ the yield of inclusive deep inelastic scattering leptons in the same $(\nu,Q^2)$-bin. For the deuterium target, i.e.~the nominator of Eq.~(\ref{eq:multiplicity-ratio}), we simply use the isospin averaged results of a proton and a neutron target. Thus in the case of deuterium we neglect the FSI of the produced hadrons and also the effect of shadowing and Fermi motion.

\subsection{No prehadronic interactions}
Before discussing the effect of prehadronic interactions we show the modifications of the multiplicity ratio due to the conventional hadronic FSI after the formation time $\tau_f$ and explore the sensitivity of the results on the size of $\tau_f$. We, therefore, neglect any interactions during $\tau_f$ and for simplicity assume that $\tau_f$ is a constant in the rest frame of each hadron independent of the particle species. Due to time dilatation the formation time $t_f$ in the laboratory frame is then proportional to the particle's energy (cf.~Eq.~(\ref{eq:lab-formation-time})).
Fig.~\ref{fig:wolead} shows the result for the $^{84}$Kr target using formation times in the range $\tau_f=0$--1.5 fm/$c$. For $\tau_f$=0 (dashed line) we get a much too strong attenuation both in the $z_h$ and $\nu$ spectrum. In this case all reaction products start to interact immediately after the $\gamma^*N$ interaction and there is no effect of time dilatation ($t_f=\tau_f=0$). As a consequence this limit leads to an almost flat $\nu$ dependence of $R_M^h$. We note that without the cut on the hadron energy $E_h$ and without the limitation by the HERMES detector acceptance one would find a strong increase of the hadron multiplicity at low $z_h$ due to particle creation in the FSI (cf.~Fig.~\ref{fig:Krdetect}). However, after applying all cuts one is left with a 10\% effect only, as can be seen from the $z_h$ spectrum in Fig.~\ref{fig:wolead}.

A slight increase of the formation time $\tau_f$ from zero to 0.1 fm/$c$ (dotted line) already leads to a dramatic change in $R_M^h$. By using this unphysically small formation time one obtains a good description of the $\nu$ dependence, but fails to reproduce the high energy part of the spectrum in $z_h$. The reason is that many of the high-energy particles, which are directly created in the primary $\gamma^*N$ interaction, now escape the nucleus due to time dilatation. Especially the formation times $t_f$ of the light pions start to exceed the dimension of the $^{84}$Kr nucleus for energies larger than $E_\pi\approx 13$ GeV. One also sees that the multiplicity ratio at low $z_h$ drops below one because both the absolute number of FSI and the energy available for particle production in the FSI is reduced. The latter decreases the probability that the secondaries survive the experimental cuts.

At larger values of $\tau_f$ one still has some attenuation due to the FSI of more massive hadrons. As seen in Fig.~\ref{fig:leading} a large fraction of the finally detected pions stem from the decay of neutral (and charged) $\rho$ mesons created in the string fragmentation. Due to their relatively large mass the latter have a smaller formation time in the lab frame and are subject to hadronic FSI, especially if they carry only a small fraction $z_h$ of the photon energy. For example, the formation length of a $\rho$ meson produced by a 7 GeV photon is always smaller than about $\gamma\cdot\tau_f\cdot c\approx9\cdot\tau_f\cdot c$. For $\tau_f=1.5$ fm/$c$ (short dashed line) only the formation lengths of the massive vector mesons and nucleons with $z_h\lesssim 0.6$--0.7 are short enough to give rise to attenuation. We mention that the small deviation from unity of $R_M^h$ at $z_h\approx 1$ is due to the Fermi motion of the bound nucleons that affects the maximum energy available for hadron production in the initial $\gamma^*N$ interaction.

\subsection{Constituent quark model}
\label{ssec:HERMES-simple}
\begin{figure}
    \includegraphics[width=12cm]{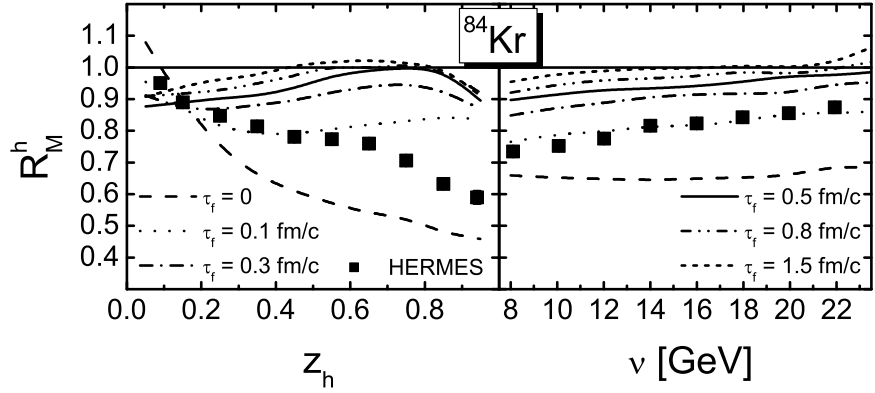}
  \caption{Calculated multiplicity ratio of charged hadrons for the $^{84}$Kr target (at HERMES) assuming no prehadronic interactions during the formation time $\tau_f=0$--1.5 fm/$c$. The data are taken from Ref.~\cite{HERMESDIS_new}.} \label{fig:wolead}
\end{figure}

From Fig.~\ref{fig:wolead} one extracts that for reasonable formation times $\tau_f\gtrsim 0.5$ fm/$c$ the (pre-) hadronic interactions have to set in quite early after the $\gamma^*N$ interaction, especially for the hadrons at large $z_h$. The Lund model predicts a vanishing prehadron production time as $z_h\to 1$ as indicated by Eq.~(\ref{eq:production-time}). However, as we have shown in Ref.~\cite{Fal03} the Lund production time of Bialas and Gyulassy \cite{Bia87} does not reproduce the charged hadron multiplicity ratios if one also accounts for the production of secondary particles in the FSI. As in our previous approach \cite{Fal03} we, therefore, set the production time of {\it all} prehadrons to zero, but now rescale the prehadronic cross sections according to the constituent quark concept (\ref{eq:prehadrons}).

\begin{figure}
    \includegraphics[width=12cm]{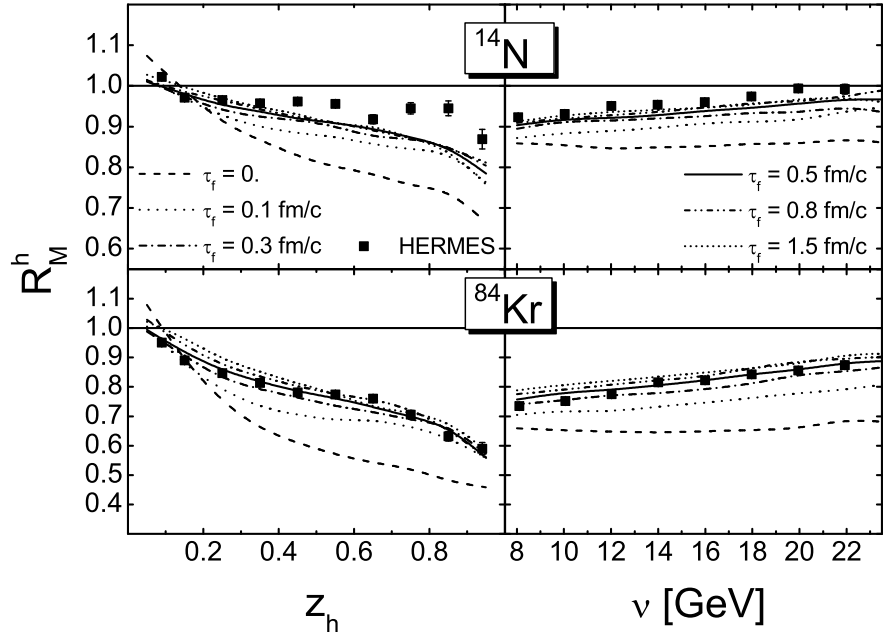}
  \caption{Calculated multiplicity ratio of charged hadrons for $^{14}$N and $^{84}$Kr nuclei (at HERMES) using the constituent quark model (\ref{eq:prehadrons}) for the prehadronic cross section and different values of the formation time $\tau_f=0$--1.5 fm/$c$. The data are taken from Ref.~\cite{HERMESDIS_new}.}
   \label{fig:NKrhadrons}
\end{figure}

Fig.~\ref{fig:NKrhadrons} shows the results of our simulation for $^{14}$N and $^{84}$Kr using formation times $\tau_f=0$--1.5 fm/$c$. The dashed lines ($\tau_f=0$) coincide with the ones in Fig.~\ref{fig:wolead} since they only involve hadronic FSI. By comparing the $\nu$ spectra of Figs.~\ref{fig:NKrhadrons} and \ref{fig:wolead} for finite formation times one observes that $R_M^h(\nu)$ is reduced by the prehadronic FSI, which also improve the agreement of $R_M^h(z_h)$ with the experimental $^{84}$Kr data at large $z_h$. However, the attenuation of the $z_h$ spectra is too strong for the $^{14}$N target which might already indicate a deficiency of our simple prehadron concept. The experimental $^{84}$Kr data favor formation times $\tau_f\gtrsim$0.3 fm/$c$ with only little sensitivity for larger values because of the finite size of the $^{84}$Kr nucleus. These times are in line with our simple estimate via the hadronic radius and with the values $\tau_f= 0.4$--0.8 fm/$c$ \cite{Cas02} extracted from antiproton attenuation in $pA$ reactions at AGS energies of 12.3 GeV and 17.5 GeV on various nuclear targets \cite{E910}. A cleaner discrimination between $\tau_f=0.3$ fm/$c$ and larger formation times will be possible in an experimental investigation at Jefferson Lab \cite{JLab} with heavier targets and lower photon energies.

\begin{figure}
    \includegraphics[width=8cm]{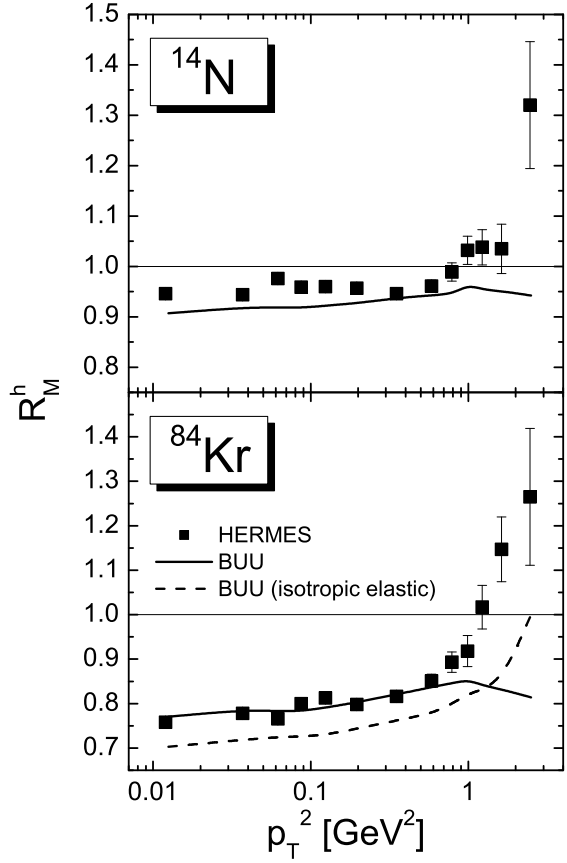}
  \caption{The multiplicity ratio of charged hadrons for $^{14}$N and $^{84}$Kr (at HERMES) as a function of the transverse momentum squared $p_T^2$. In the simulation we use the constituent quark concept (\ref{eq:prehadrons}) for the prehadronic cross sections and a formation time $\tau_f=0.5$ fm/$c$. In the simulation indicated by the dashed line we additionally assume that all elastic scatterings are isotropic in their center-of-mass frame. The data are taken from Ref.~\cite{HERMESDIS_new}.}
   \label{fig:NKrpt}
\end{figure}

In Fig.~\ref{fig:NKrpt} we show the transverse momentum dependence of the multiplicity ratio (\ref{eq:multiplicity-ratio}), where the transverse component $p_T$ of the hadron is defined with respect to the momentum direction of the virtual photon. In the simulation we use the constituent quark concept (\ref{eq:prehadrons}) for the prehadronic cross sections and the formation time $\tau_f=0.5$ fm/$c$. We expect that the $p_T$ distribution of the observed hadrons is broadened for complex nuclei compared to deuterium due to multiple scattering of the (pre-)hadrons. Up to $p_T^2\approx 1$ GeV$^2$ the $p_T^2$ dependence of the multiplicity ratio is well reproduced for both the nitrogen and the krypton target. However, the data of Ref.~\cite{HERMESDIS_new} show a strong increase of $R_M^h$ for $p_T^2\gtrsim 1$ GeV$^2$, which is not reproduced by our (pre-)hadronic FSI even if one assumes that all elastic scattering events are isotropic in the center-of-mass system (dashed line).  This can be considered a signal for a partonic origin of the enhancement of high-$p_T$ hadrons in $eA$ collisions either via a change of the parton distributions inside the nuclear medium and/or the Cronin effect \cite{Cro75,Kop02,Acc04}.

The Cronin effect was first observed in 1975 \cite{Cro75} via an enhancement of high-$p_T$ hadrons in $pA$ collisions and has become especially important recently in connection with data from high-$p_T$ hadron production in heavy-ion collisions \cite{RHIC,Cass04}. Similar to $pA$ collisions \cite{Kop02}, a high-energy parton produced in a direct $\gamma^*N$ interaction may be subject to soft coherent and incoherent multiple rescatterings in the nuclear medium. While the incoherent rescatterings can be interpreted intuitively as a random walk in transverse momentum space \cite{Joh01}, the coherent gluon radiation from different nucleons is subject to Landau-Pomeranchuk-Migdal interference effects \cite{Wang}. The authors of Ref.~\cite{Kop03} have calculated the transverse momentum broadening that is caused by the multiple scattering of the struck parton before the prehadron production time. Their calculation reproduces the strong increase of $R_M^h(p_T)$ for $p_T>1$ GeV/$c$ and thus supports the interpretation in terms of a Cronin effect.

In our simulation we do not find any dependence of the charged hadron multiplicity ratio (\ref{eq:multiplicity-ratio}) on the photon virtuality $Q^2$. This is no surprise since in the ratio $R_M^h$ the $Q^2$ dependence of the primary electroproduction cross section cancels out and our prehadronic cross sections (\ref{eq:prehadrons}) do not depend on the virtuality of the photon. Therefore, an experimentally observed enhancement of $R_M^h$ with $Q^2$ could be interpreted as a signature for color transparency \cite{CT,Kop01}. We will show in Sec.~\ref{sec:EMC} that indeed the simple constituent quark ansatz for the prehadronic cross sections (\ref{eq:prehadrons}) overestimates the attenuation in the kinematic regime of the EMC experiment, i.e.~at larger values of $\nu$ and $Q^2$.

\subsection{Acceptance cuts}
\begin{figure}
    \includegraphics[width=12cm]{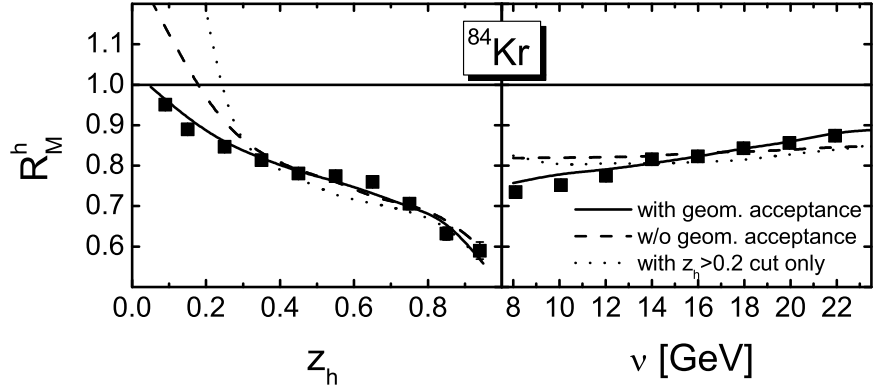}
  \caption{Influence of the HERMES detector geometry on the observed multiplicity ratios for a $^{84}$Kr target. In the simulation we use the constituent quark concept (\ref{eq:prehadrons}) for the prehadronic cross sections and a formation time $\tau_f=0.5$ fm/$c$. The solid line represents the result of our simulation when accounting for the geometrical acceptance of the HERMES detector. In the simulation indicated by the dashed line no acceptance cuts have been employed. The dotted line represents the result of a simulation where in addition to the detector acceptance the $E_h>1.4$ GeV cut has been neglected. The data are taken from Ref.~\cite{HERMESDIS_new}.}
   \label{fig:Krdetect}
\end{figure}

We now discuss how the geometrical acceptance of the HERMES detector and the kinematic cuts affect the multiplicity ratio (\ref{eq:multiplicity-ratio}). In Fig.~\ref{fig:Krdetect} we compare the results of our simulation using the constituent quark concept (\ref{eq:prehadrons}) and a formation time $\tau_f=0.5$ fm/$c$ for the HERMES acceptance (solid line) in comparison to a $4\pi$-detector (dashed line). In both calculations we still account for all kinematic cuts in the HERMES experiment. As can be seen from the $z_h$ spectrum, a detector with full angular coverage (dashed line) will detect many more of the low-energy particles -- produced in the FSI -- which simply do not end up in the HERMES detector. As a result, the $\nu$ spectrum for a $4\pi$-detector is almost flat since an increase of the formation time with $\nu$ due to time dilatation not only reduces the attenuation but also the particle production in the FSI. According to our simulations the slope in the $\nu$ spectrum experimentally observed at HERMES partly arises because at lower photon energies particle production in the FSI is less forward peaked and, therefore, less particles are seen by the HERMES detector. Note, that this problem is usually neglected in other approaches \cite{Wang,Arl03,Acc02,Kop03} that intend to describe the observed multiplicity ratios. The dotted line in Fig.~\ref{fig:Krdetect} shows the result of a simulation where in addition to the geometrical acceptance the $E_h>1.4$ GeV cut has been neglected. As can be seen from the low $z_h$ part of the multiplicity ratio, one now detects even more low-$z_h$ hadrons. Without the cut $z_h>0.2$ the multiplicity ratio as a function of the photon energy rises to $R_M^h(\nu)\approx 1.5$. 

Summarizing this paragraph we stress that one has to take the geometrical acceptance of the HERMES detector into account if one wants to draw conclusions from a comparison of experimental with theoretical results for the low-$z_h$ part and the $\nu$ dependence of the multiplicity ratio.

\subsection{Double hadron attenuation}

Before we turn to the individual attenuation of the various identified hadrons we compare our simulation with the recently measured double-hadron attenuation at HERMES \cite{Nezza}. In each event only the two (charged {\it or} neutral) hadrons with the highest energies are considered. In the following we denote the hadron with the highest $z_h$ as the leading hadron and the other one as the subleading hadron. The experimental observable is the double-hadron attenuation ratio
\begin{equation}
\label{eq:double-hadron}
    R_2(z_2)=\frac{\frac{N_2(z_2)}{N_1}\big|_A}{\frac{N_2(z_2)}{N_1}\big|_D}.
\end{equation}
Here $N_2(z_2)$ denotes the number of events where the leading and subleading hadron carry the energy fraction $z_1>0.5$ and $z_2<z_1$, respectively, and $N_1$ is the number of events where at least one of them has $z_h>0.5$. The kinematic cuts are the same as for charged hadrons except for the Bjorken variable $x$, which now has the new boundary $x>0.01$.

\begin{figure}
    \includegraphics[width=10cm]{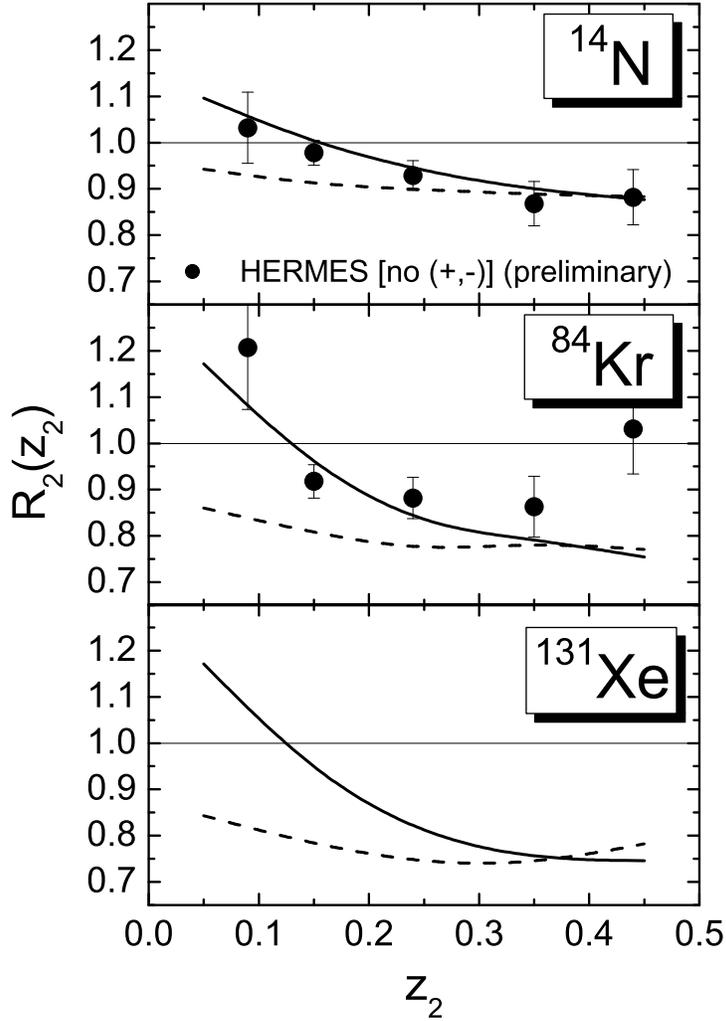}
  \caption{Double-hadron attenuation ratio $R_2$ for a $^{14}$N, $^{84}$Kr and $^{131}$Xe target as a function of the energy fraction $z_2$ of the subleading hadron. In the simulation (solid line) we use the constituent quark concept (\ref{eq:prehadrons}) for the prehadronic cross sections and a formation time $\tau_f=0.5$ fm/$c$. To exclude contributions from $\rho^0$ decay into $\pi^+\pi^-$ the charge combinations '$+-$' and '$-+$' have been excluded both in experiment and in the simulation. The dashed line shows a calculation with a purely absorptive treatment of the FSI. The preliminary HERMES data are taken from Ref.~\cite{Nezza}.}
   \label{fig:double}
\end{figure}

Fig.~\ref{fig:double} shows the double-hadron multiplicity ratio (\ref{eq:double-hadron}) for $^{14}$N, $^{84}$Kr and $^{131}$Xe. To exclude contributions from $\rho^0$ decay into $\pi^+\pi^-$ the charge combinations '$+-$' and '$-+$' have been excluded both in experiment and in the simulation. The solid line shows the result of a full coupled-channel calculation using the constituent quark concept (\ref{eq:prehadrons}) and the formation time $\tau_f=0.5$ fm/$c$. The shape of the spectrum is similar to that of $R_M^h(z_h)$ of the charged hadron multiplicity ratio shown in Fig.~\ref{fig:NKrhadrons}. The reason is quite simple: For the interpretation we discard for a moment the constant factors $N_1$ in Eq.~(\ref{eq:double-hadron}) and the factors $N_e$ in Eq.~(\ref{eq:multiplicity-ratio}) which have no influence on the shape of the $z_h$ dependence. The only difference between $N_h(z_h)|_A/N_h(z_h)|_D$ and $N_2(z_2)|_A/N_2(z_2)|_D$ then is that one restricts the detected hadron to the subleading particle in the latter case. If the subleading particle (with energy fraction $z_2$) of the initial $\gamma^*N$ reaction interacts with the nuclear environment, it will produce a bunch of low-energy particles. The {\it new} subleading hadron in the event then has a energy fraction $z_2'<z_2$. As for the usual charged hadron multiplicity spectrum the coupled-channel FSI shuffle strength from the high $z_h$ part to the low $z_h$ part of the spectrum. This is not the case for purely absorptive FSI (dashed line in Fig.~\ref{fig:double}). As one can see, our coupled-channel calculations (solid line) -- using the constituent quark concept (\ref{eq:prehadrons}) and the formation time $\tau_f=0.5$ fm/$c$ -- are again in quantitative agreement with the experimental data apart from the last data point in the $^{84}$Kr data, which indicates a multiplicity ratio $R_2(z_2=0.5)\approx 1$. This behavior cannot be explained within our model.

\begin{figure}
    \includegraphics[width=12cm]{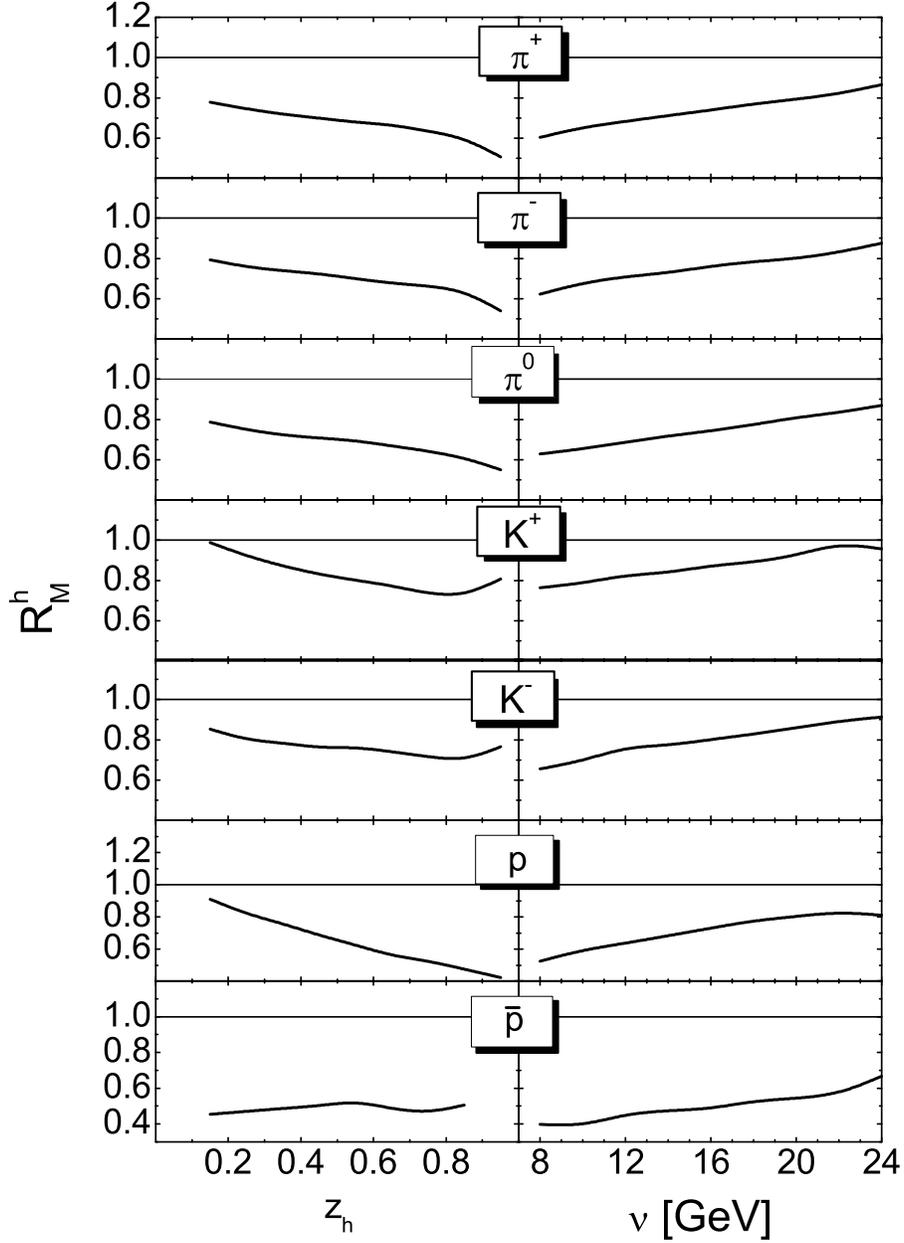}
  \caption{Predictions for the multiplicity ratios of $\pi^{\pm}$, $\pi^0$, $K^\pm$, $p$ and $\bar{p}$ for a $^{131}$Xe nucleus (at HERMES) as a function of the hadron energy fraction $z_h$ and the photon energy $\nu$. In the simulation we use the constituent quark concept (\ref{eq:prehadrons}) for the prehadronic cross sections and a formation time $\tau_f=0.5$ fm/$c$.} \label{fig:Xeid}
\end{figure}

As can be seen from Fig.~\ref{fig:double}, our calculations predict about the same double-hadron attenuation ratio for $^{131}$Xe and $^{84}$Kr. The reason is that the attenuation of leading and double hadrons increases in the same way when going from the krypton to the xenon target. Hence, the double-hadron attenuation ratio (\ref{eq:double-hadron}) stays roughly the same. Note, that this does not necessarily imply the same hadron attenuation for $^{84}$Kr and $^{131}$Xe. As we have shown in Ref.~\cite{Fal03}, and as one can also see by comparing the multiplicity ratios for the two targets in Figs.~\ref{fig:Krid} and \ref{fig:Xeid}, the hadron attenuation in the $^{131}$Xe nucleus is on average 5\% larger than for $^{84}$Kr.

\subsection{Attenuation of identified hadrons}
We finally consider the attenuation of $\pi^\pm$, $\pi^0$, $K^\pm$, $p$ and $\bar{p}$ in DIS of 27.6 GeV positrons off $^{20}$Ne, $^{84}$Kr and $^{131}$Xe nuclei. For the krypton and xenon target the cuts are the same as for charged hadrons plus the momentum cuts necessary for particle identification at HERMES \cite{HERMESDIS_new}, i.e.~2.5 GeV/$c$ $<p_{\pi,K}<15$ GeV/$c$ and 4 GeV/$c$ $<p_{p,\bar{p}}<15$ GeV/$c$. The following cuts are different for the neon target \cite{Elb03}: 2 GeV$<\nu<24$ GeV, 0.6 GeV/$c$ $<p_\pi<15$ GeV/$c$ and 2 GeV/$c$ $<p_K<15$ GeV/$c$.

\begin{figure}
    \includegraphics[width=12cm]{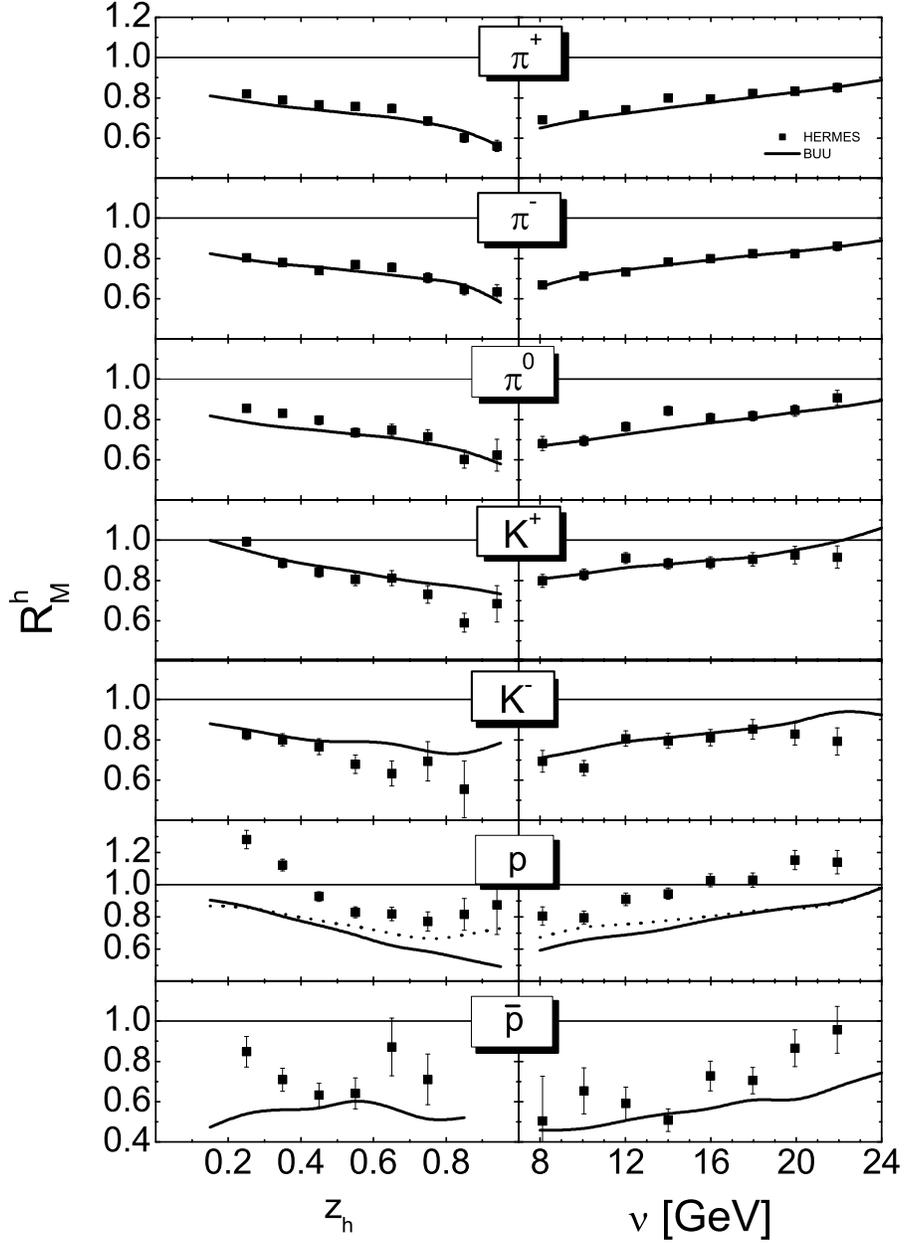}
  \caption{Multiplicity ratios of $\pi^{\pm,0}$, $K^\pm$, $p$ and $\bar{p}$ for a $^{84}$Kr nucleus (at HERMES) as a function of the hadron energy fraction $z_h=E_h/\nu$ and the photon energy $\nu$. The solid line represents the result of a simulation, where we use the constituent quark concept (\ref{eq:prehadrons}) for the prehadronic cross sections and a formation time $\tau_f=0.5$ fm/$c$. The dotted line in the proton spectrum indicates the result of a simulation where all $\gamma^*N$ events are created by PYTHIA. The data are taken from Ref.~\cite{HERMESDIS_new}.}
  \label{fig:Krid}
\end{figure}

\begin{figure}
    \includegraphics[width=12cm]{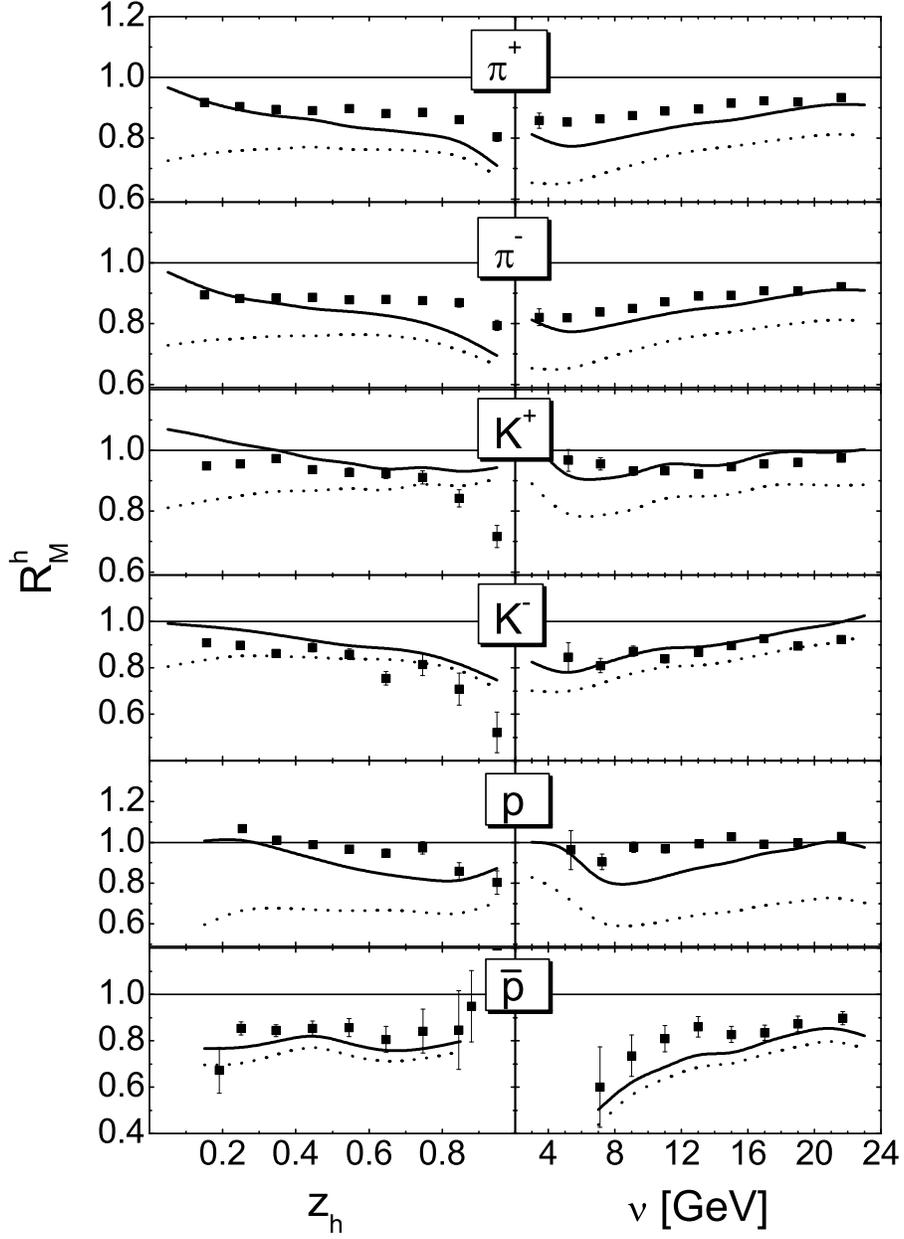}
  \caption{Multiplicity ratios of $\pi^{\pm}$, $K^\pm$, $p$ and $\bar{p}$ for a $^{20}$Ne nucleus (at HERMES) as a function of the hadron energy fraction $z_h$ and the photon energy $\nu$. The solid line represents the result of a simulation where we use the constituent quark concept (\ref{eq:prehadrons}) for the prehadronic cross sections and a formation time $\tau_f=0.5$ fm/$c$. The dotted line represent the result of a simulation with a purely absorptive treatment of the FSI. The data are taken from Ref.~\cite{Elb03}.} 
  \label{fig:Neid}
\end{figure}

For the moment we stick again to the prehadron concept (\ref{eq:prehadrons}) and the formation time $\tau_f=0.5$ fm/$c$ in our simulation. Figs.~\ref{fig:Krid} and \ref{fig:Neid} show the multiplicity ratios on $^{84}$Kr and $^{20}$Ne, respectively. The attenuation of pions is well described for the $^{84}$Kr nucleus while it is slightly too strong for the light $^{20}$Ne target. This is no surprise since for $^{84}$Kr our model already reproduced the multiplicity ratio of charged hadrons -- dominated by pions -- while the attenuation was too strong in the case of the lighter $^{14}$N nucleus.

\begin{figure}
    \includegraphics[width=14cm]{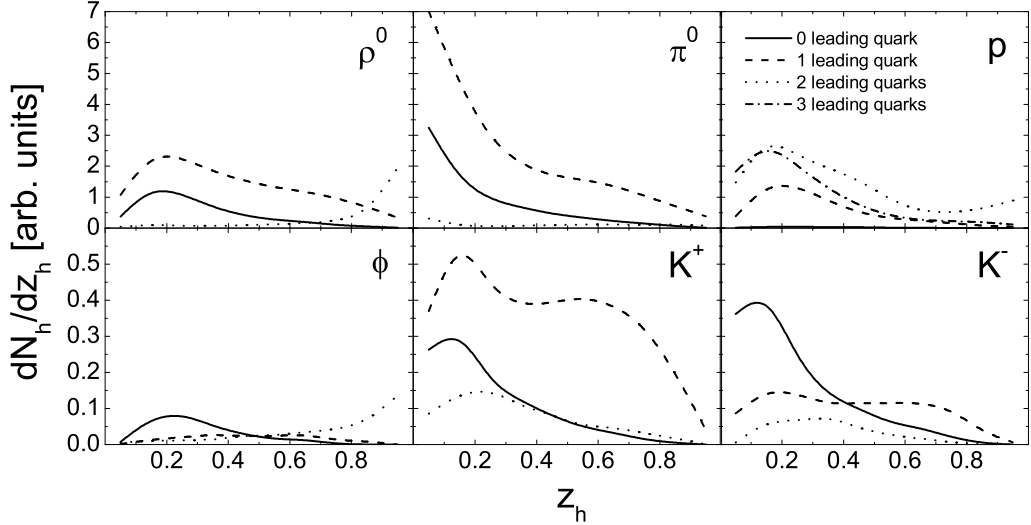}
  \caption{Energy spectra of hadrons produced in electron-nucleon interactions at HERMES. The different lines indicate the hadrons that contain zero (solid line), one (dashed line), two (dotted line) or three quarks (dash-dotted line) from the beam or target. For the proton the solid line nearly coincides with the $z_h$ axis, which implies that most protons contain at least one leading quark.}
   \label{fig:leading}
\end{figure}

The attenuation of $K^+$ mesons is well described for the heavier krypton target while it is in poor agreement with the $^{20}$Ne data. In the latter case the calculated multiplicity ratio is larger than one at small $z_h$ in contradiction to the experimental data. For both nuclei our simulation yields approximately the same attenuation for $K^-$ and $K^+$ mesons at large $z_h$. The reason for this is directly related to the $K^+$ and $K^-$ spectra in Fig.~\ref{fig:leading}, which show that due to the quark content ($\bar{u}s$) the $K^-$ contain less quarks from the beam or target than the $K^+$. The few $K^-$ that are not solely made of quarks and antiquarks created from the vacuum in the string fragmentation carry (anti-)quarks from the resolved photon component or the nucleon sea. According to the constituent quark concept (\ref{eq:prehadrons}) one has thus less prehadronic interactions of $K^-$. This is compensated by the larger $K^-$-nucleon cross section. In total this leads to a similar attenuation of $K^+$ and $K^-$. From Fig.~\ref{fig:leading} it can also be seen that a large part of the kaons at high $z_h$ stem from $\phi$ decay into $K^+K^-$. The attenuation of $K^\pm$ at high $z_h$, therefore, strongly depends on the FSI of the $\phi$ meson; this is in analogy to the attenuation of charged hadrons (pions) which are strongly affected by the FSI of the $\rho$ meson.

There is a further complication connected with the multiplicity ratio of kaons, which is neglected in a purely absorptive treatment of the FSI. The initial $\gamma^*N$ interaction produces many more pions and $\rho$ mesons than strange particles (cf.~Fig.~\ref{fig:leading}). These high-energy particles can produce secondary $K^+$ and $K^-$ in the nuclear FSI and thereby enhance the multiplicity ratio for $K^\pm$ at low $z_h$. This is illustrated by the dotted lines in Fig.~\ref{fig:Neid} which shows the result of a purely absorptive treatment of the FSI, in which every particle that undergoes FSI is simply removed from the simulation. One clearly sees that kaon absorption in the FSI is compensated to a large extent by the production of kaons in the nuclear FSI of pions and $\rho$ mesons. For $z_h\lesssim$0.35 the $K^+$ production exceeds the absorption in the light $^{20}$Ne and leads to a multiplicity ratio larger than one. This is not the case for the $^{84}$Kr nucleus which is large enough to also absorb some of the secondary kaons. Of course this effect strongly depends on the strangeness production cross section used in the FSI. Unless one does not have all these coupled channels under control it is, therefore, hard to draw any conclusion from the kaon attenuation in DIS off nuclei.

From Fig.~\ref{fig:leading} one also sees that there are only very few protons that contain no quarks from the beam or target because diquark-antidiquark creation is strongly suppressed in the string fragmentation due to the relatively large diquark masses. The latter also explains why most protons contain two of the original quarks, i.e.~the diquark from the target nucleon. These remnants have a very large prehadronic cross section according to the constituent quark concept (\ref{eq:prehadrons}). As a result the proton multiplicity ratio comes out too small for both the $^{84}$Kr and the $^{20}$Ne nucleus. As discussed in Sec.~\ref{ssec:Lund-prehadrons}, the Lund model points to a larger formation time for protons than for the lighter pions, $\rho$ mesons and kaons (cf.~Eq.~(\ref{eq:formation-time})). However, as we have shown in Fig.~\ref{fig:NKrhadrons}, a further increase of the formation time to $\tau>0.5$ fm/$c$ does not change our result at large $z_h$ as long as the production time is zero. Since proton attenuation at large $z_h$ is solely due to prehadronic interactions this either points towards a smaller prebaryonic cross section or a finite production time.

A further problem connected with the proton spectra is the strong increase of the multiplicity ratio for $z_h<0.4$ which is seen in the experimental $^{84}$Kr data. At small proton energies $R_M^p$ becomes larger than one which might be understood in our model by a slowing down of high-energy protons in the FSI. Alternatively, protons might be knocked out of the nucleus in the FSI of a high-energy meson produced in the primary $\gamma^*N$ interaction. We do indeed see these effects in our simulation, however, in both cases the experimental momentum cut $p_p>4$ GeV/$c$ removes most of these protons from the acceptance. Thus, the protons in our transport simulation loose too much energy per collision in the (pre-)hadronic FSI scenario.

As we have pointed out in the discussion of Fig.~\ref{fig:elementary} the use of FRITIOF in the simulation of $\gamma^*N$ events has a small effect on the proton spectra. In Fig.~\ref{fig:Krid} we, therefore, also show the proton attenuation in a simulation where {\it all} $\gamma^*N$ interactions above $W_\mathrm{PY}=3$ GeV are simulated by PYTHIA. One observes that besides a slight improvement at large $z_h$ and low photon energies $\nu$ this has no effect on our result.

Within our simulation the attenuation of antiprotons also comes out slightly too large in the (pre-)hadronic FSI scenario. We find that the antiprotons with $z_h\gtrsim0.5$ are mainly beam or target remnants that contain an antiquark from the resolved photon or the nucleon sea, whereas most of the antiprotons with $z_h\lesssim 0.5$ are solely made of antiquarks that are produced in the fragmentation of the string excited in the $\gamma^*N$ interaction. According to the constituent quark concept (\ref{eq:prehadrons}) the attenuation of antiprotons with $z_h<0.5$ is, therefore, only caused by hadronic FSI after $\tau_f$.

In Fig.~\ref{fig:Xeid} we also show predictions for the attenuation of identified hadrons on a $^{131}$Xe target. In the simulation we have used the constituent quark concept (\ref{eq:prehadrons}) and a formation time $\tau_f=0.5$ fm/$c$. By comparing the multiplicity ratios $R_M^h$ of negatively and positively charged pions for $^{20}$Ne, $^{84}$Kr, and $^{131}$Xe nuclei, we find that the attenuation ($1-R_M^h$) scales like $A^{\alpha}$ with an exponent $\alpha=0.22$--0.29 at $z_h=0.95$. If the attenuation was simply proportional to the distance that the particles propagate through the medium, one would naively expect a scaling exponent $\alpha=1/3$. The deviation is caused by the finite formation time and the nonuniform density distribution in nuclei. At lower values of $z_h$ or in the integrated $\nu$ spectra the scaling behavior is hidden by the coupled channel effects. We note that a scaling of the attenuation with the target mass $\sim(A)^{2/3}$, as predicted by Ref.~\cite{Wang}, would imply an increase of about 34\% when using $^{131}$Xe instead of $^{84}$Kr.

\subsection{Lund production and formation times}
\label{ssec:HERMES-Lund}

We now test the result of a simulation where both the production times of the prehadrons and the formation time $\tau_f$ of each individual particle are explicitly extracted from the corresponding string fragmentation in JETSET (cf.~Fig.~\ref{fig:JETSET2}). As described in Sec.~\ref{ssec:Lund-prehadrons} there are in principle three time scales involved in the Lund fragmentation process: i) The production proper time $\tau_{p1}$ of the hadron's first constituent, which is obviously zero if the hadron contains a constituent from a string end, ii) the production proper time $\tau_{p2}$ when the second constituent is produced and a color neutral object is formed, and iii) the formation proper time $\tau_f$ where the two world lines of the constituents cross for the first time. 

We point out that -- due to technical reasons -- all particles that emerge from the primary $\gamma^*N$ interaction start to propagate from the interaction vertex, while the production time only affects the beginning of their interactions with the nuclear medium. One may therefore expect slight deviations from the real reaction geometry, where the excited string propagates over a small distance prior to fragmentation. 

\begin{figure}
    \includegraphics[width=12cm]{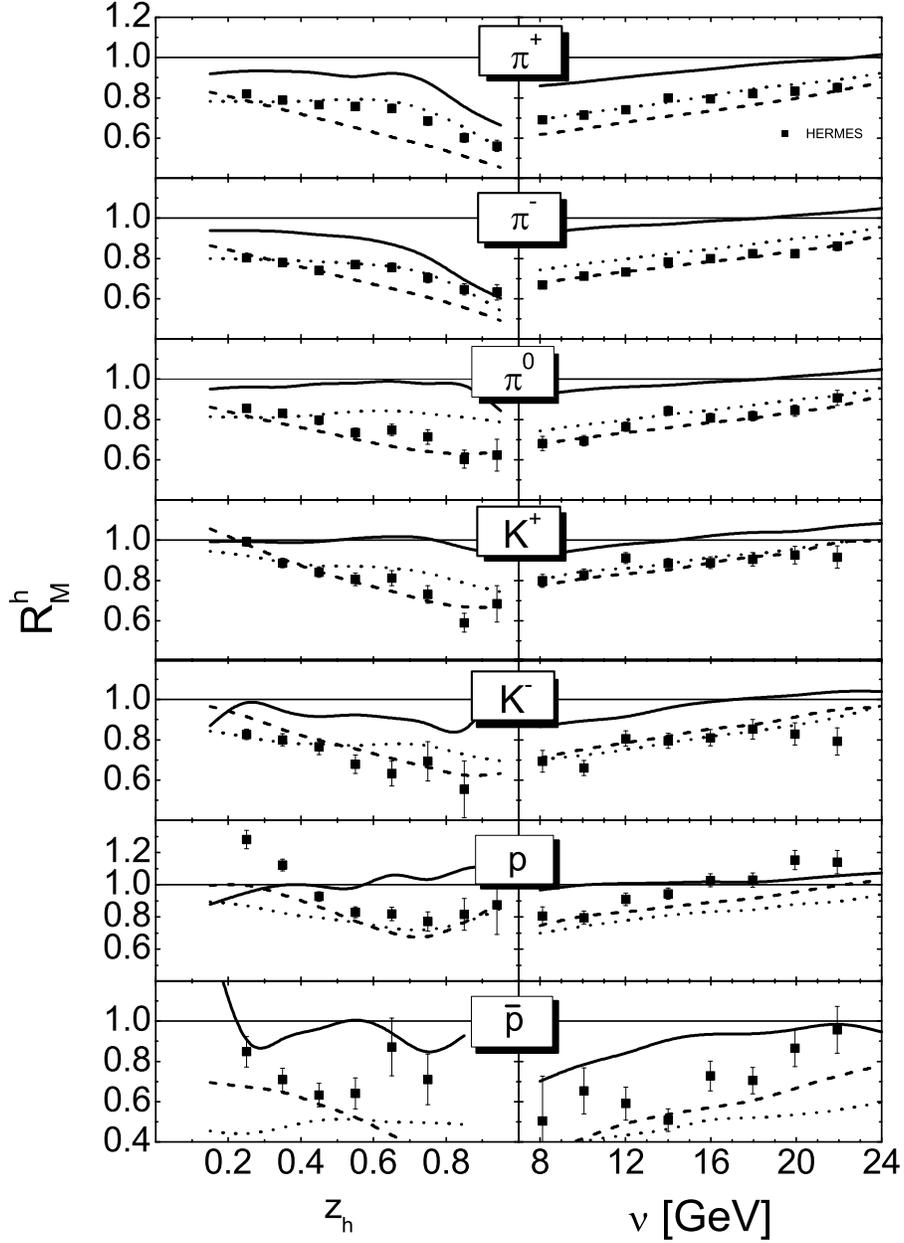}
  \caption{Multiplicity ratios of $\pi^{\pm}$, $\pi^0$, $K^\pm$, $p$ and $\bar{p}$ for a $^{84}$Kr nucleus (at HERMES) as a function of the hadron energy fraction $z_h$ and the photon energy $\nu$. In the simulation we use the proper times $\tau_{p_2}$ (solid line), $0.2\tau_{p_2}$ (dotted line) and $\tau_{p_1}$ (dashed line) from the JETSET routine as the prehadron production time. The prehadronic cross section is set to the full hadronic cross section and interactions before the production time are neglected.}
\label{fig:KridJETSET}
\end{figure}

We now proceed and introduce two effective cross sections: $\sigma_1$ which accounts for the 'partonic' interaction between $\tau_{p1}$ and $\tau_{p2}$, and $\sigma_2$ which accounts for the 'prehadronic' interactions of the color neutral object between $\tau_{p2}$ and $\tau_f$. We first neglect the 'partonic' interactions ($\sigma_1=0$) and set the 'prehadronic' cross section $\sigma_2$ equal to the full hadronic cross section $\sigma_h$. In such a scenario we are no longer sensitive to the formation time $\tau_f$ but only on $\tau_{p2}$. The solid lines in Fig.~\ref{fig:KridJETSET} show the result of such a simulation for a $^{84}$Kr target. Since the resulting attenuation is much too weak we conclude that the strong FSI have to start earlier. Note that a reduction of $\sigma_2$ will further enhance the discrepancy with experimental data. To achieve reasonable results for the multiplicity ratio one has to rescale $\tau_{p2}$ by about a factor of 0.2 (dotted line), which is quite dramatic. According to the Lund model (cf.~Eq.~(\ref{eq:production-time})) this has to be interpreted as an {\it increase} of the string tension $\kappa$ by an unreasonably large factor of about {\it five} in the nuclear medium. Setting $\sigma_1$ to zero implies to neglect any interaction of the nucleon debris with the nuclear medium between the moment of the $\gamma^*N$ interaction and $\tau_{p2}$. This might be a problem, since the hadronic string that is produced in the DIS may interact with a hadronic cross section right from the beginning \cite{Cio02}.

The average size of the prehadron-production times of the gluon-bremsstrahlung model~\cite{Kop03} is about a factor of ten smaller than the times $t_{p2}$ extracted from JETSET. In fact, their average size is rather compatible with $t_{p1}$. If one assumes that strong FSI already set in right after $\tau_{p1}$ and sets $\sigma_1=\sigma_2=\sigma_h$ one gets the result indicated by the dashed curves in Fig.~\ref{fig:KridJETSET}, which are in satisfactory agreement with the experimental data. However, such a large interaction cross section $\sigma_1$ is definitely not of perturbative nature. Furthermore, this recipe again implies that all beam and target remnants can interact right after the photon-nucleon interaction since the production time $\tau_{p1}$ of their first constituent is zero. In fact, this scenario is not much different from our constituent quark ansatz discussed before. Again the interactions of the beam and target remnants may be considered as effectively accounting for the interaction of the strings right after their creation in the $\gamma^*N$ interaction. The solid squares in Fig.~\ref{fig:JETSET2} indicate the average starting times of the prehadronic and hadronic interactions according to the constituent quark model (\ref{eq:prehadrons}) with $\tau_f=0.5$. The $z_h$ dependence of these two times has to be compared with $t_{p1}$ and $t_f$, respectively. In both cases the shape looks quite similar while the average times of our constituent quark concept are somewhat smaller. The latter is partly compensated by the reduced prehadronic cross section, cf.~Eq.~(\ref{eq:prehadrons}). Due to time dilatation the production times $t_{p1}$ are in general already of the order of the nuclear radius. This explains why the beam and target remnants -- for which $t_{p1}=0$ -- dominate the shape of the spectra.

\section{Hadron attenuation at EMC energies}
\label{sec:EMC}
In this section we test different space-time pictures of hadronization in comparison to the EMC data with 100 and 200 GeV muon beams \cite{EMC}. In the previous section we have seen that almost all of the HERMES data can be described with the simple prehadron concept of Sec.~\ref{ssec:transport-prehadrons}, i.e.~setting the production time $\tau_p$ to zero for all hadrons and using the constituent quark concept (\ref{eq:prehadrons}) for the prehadronic cross sections during the formation time $\tau_f$. Obviously, this picture can only represent a rough approximation to the real hadronization process. Neither is it very likely that the string fragments convert {\it instantaneously} into color neutral prehadrons nor do the cross sections {\it instantaneously} jump from the rescaled values (\ref{eq:prehadrons}) to the full hadronic size.

The kinematic regime of the EMC experiment, which uses a 100 GeV (200 GeV) muon beam, is different from that of the HERMES experiment. Here, the kinematic cuts are $Q^2>2$ GeV$^2$, $W>4$ GeV, $x>0.02$, 10 GeV$<\nu<85$ GeV (30 GeV$<\nu<170$ GeV) and $E_h>3$ GeV. In addition we again account for the angular acceptance of detector, i.e.~$\pm 5^\circ$ horizontally and $\pm 8^\circ$ vertically \cite{EMC81}.

\begin{figure}
    \includegraphics[width=8cm]{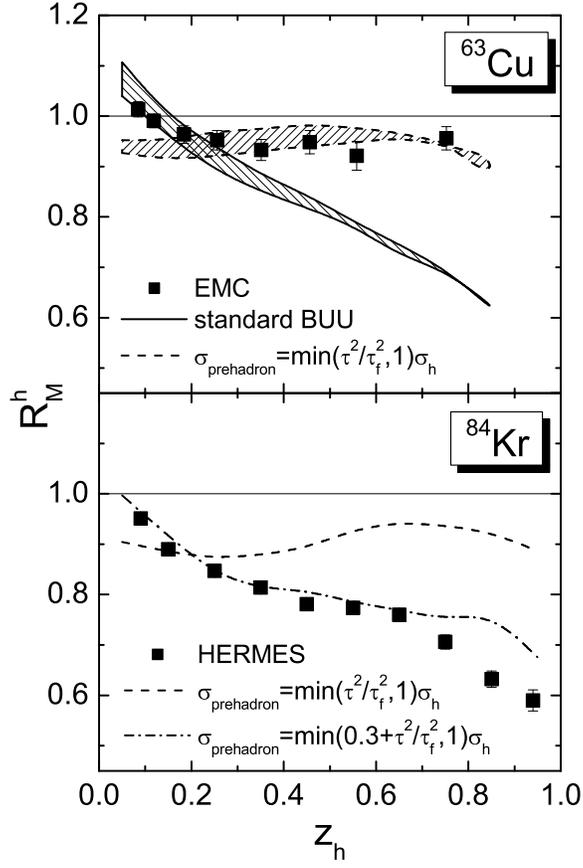}
  \caption{{\it Upper panel:} Multiplicity ratio of charged hadrons for a $^{63}$Cu target as a function of the fractional energy $z_h$ in the kinematic regime of the EMC experiment. The shaded areas are bounded by simulations using a 100 GeV (lower boundary) and 200 GeV (upper boundary) muon beam. The data are taken from Ref.~\cite{EMC}. {\it Lower panel:} Multiplicity ratio of charged hadrons for a $^{84}$Kr target as a function of the fractional energy $z_h$ in the kinematic regime of the HERMES experiment. The data are taken from Ref.~\cite{HERMESDIS_new}. The solid line shows the result of our simulation using the constituent quark concept (\ref{eq:prehadrons}) for the prehadronic cross sections and a formation time $\tau_f=0.5$ fm/$c$. In the simulation represented by the dashed line we assumed a prehadronic cross section increasing quadratically in proper time during $\tau_f=0.5$ fm/$c$ from zero up to the full hadronic size. The dash-dotted line represents a simulation where the prehadronic cross section increases quadratically in proper time from $0.3\sigma_h$ to the full hadronic cross section.}
\label{fig:emc}
\end{figure}

In the upper panel of Fig.~\ref{fig:emc} we show the EMC result for the multiplicity ratio of charged hadrons on a $^{63}$Cu target as a function of the hadron energy fraction $z_h$. The solid line shows the result for the simple prehadron concept (\ref{eq:prehadrons}) with the constant formation time $\tau_f$. The shaded area indicates the region between the results of calculations using 100 GeV and 200 GeV muon beam energy. Obviously, we get a much too strong attenuation for $z_h>0.3$. This either implies that the prehadronic interactions set in too early or that the cross sections (\ref{eq:prehadrons}) are too large.

The dashed line shows the result for a simulation when assuming that during $\tau_f$ the prehadron cross section increases quadratically in proper time $\tau$ from zero to the asymptotic value $\sigma_h$. Such an ansatz can be motivated by color transparency, which states that the cross sections of the color neutral prehadron scales with its diameter squared. While giving the right attenuation for EMC energies, such an ansatz fails to explain the HERMES data as long as the initial cross section is exactly zero (see lower panel of Fig.~\ref{fig:emc}). For the calculation indicated by the dash-dotted line in the lower panel of Fig.~\ref{fig:emc} we assumed an initial prehadronic cross section $0.3\sigma_h$ followed by a quadratic increase in proper time up to the full hadronic value $\sigma_h$. The difference between the initial prehadronic cross sections at HERMES and EMC might be explained if one assumes that the cross section right after the $\gamma^*N$ interaction is set by the resolution of the virtual photon, i.e.~$Q^2$. Indeed, the average values of $Q^2$ in the EMC experiment are more than twice as large as at HERMES energies. However, the HERMES data \cite{HERMES-web} do not indicate a strong enough $Q^2$ dependence of the multiplicity ratio to explain such a dramatic difference between the two 'initial' values for the prehadron cross section.

\begin{figure}
    \includegraphics[width=10cm]{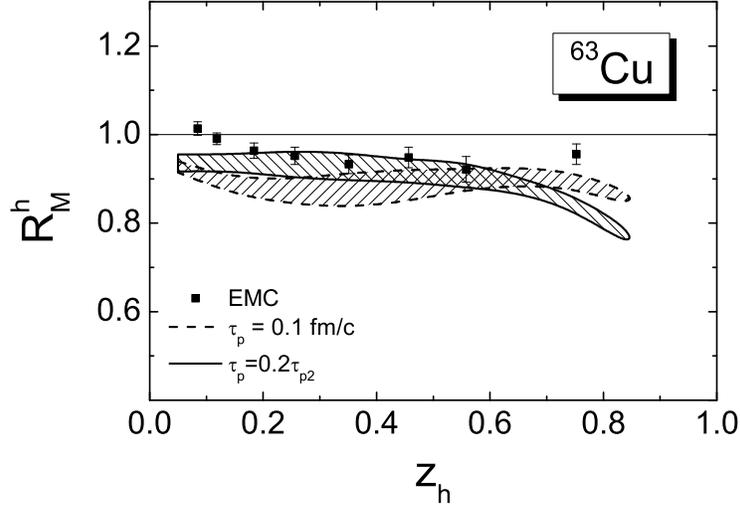}
  \caption{Multiplicity ratio of charged hadrons for a $^{63}$Cu target as a function of the fractional energy $z_h$ in the kinematic regime of the EMC experiment. The shaded areas are bounded by simulations using a 100 GeV (lower boundary) and 200 GeV (upper boundary) muon beam. The data are taken from Ref.~\cite{EMC}. The dashed line shows the result of our simulation using a constant prehadron production time $\tau_p=0.1$ fm/$c$. In the calculation indicated by the solid line we used the Lund production time $\tau_p=0.2\tau_{p2}$, where $\tau_{p2}$ has been directly extracted from JETSET. In both calculation we set the prehadronic cross section equal to the full hadronic cross section and neglect interactions before the production time.}
\label{fig:emc2}
\end{figure}

In Sec.~\ref{sec:HERMES} we have shown that, as long as one neglects a strong partonic energy loss \cite{Wang,Arl03} right after the $\gamma^*N$ interaction, the (pre-)hadronic interactions have to set in very early (cf.~Fig.~\ref{fig:wolead}) to explain the various HERMES data. Since the Lorentz $\gamma$ factors involved in the EMC experiment are about five times larger than at HERMES energies, a finite (but small) production time will have a larger impact on the calculated multiplicity ratios. In Fig.~\ref{fig:emc2} we, therefore, also show the results of a calculation with a constant production time $\tau_p=0.1$ fm/$c$ as well as the production time $\tau_p=0.2\tau_{p2}$ with $\tau_{p2}$ extracted directly from JETSET. For simplicity we neglect all (partonic) interactions before the production time of the prehadrons and set the prehadronic cross to the full hadronic cross section $\sigma_h$. 

The result for $\tau_p=0.1$~fm/$c$ in the kinematic region of the HERMES experiment is shown in Fig.~\ref{fig:wolead} and yields a too weak attenuation at large $z_h$. An additional formation time with reduced cross sections would further enhance this discrepancy. On the other hand, the production time $\tau_p=0.1$ fm/$c$ is still too small to give the right attenuation at EMC energies as can be seen from the dashed line in Fig.~\ref{fig:emc2}.

In Sec.~\ref{ssec:HERMES-Lund} we have found that using $\tau_{p1}$ as the prehadron production time yields a satisfactory description of the HERMES data. However, using $\tau_{p1}$ as the prehadron production time leads to the same problem observed for our constituent quark concept in Fig.~\ref{fig:emc}, i.e.~a too strong attenuation of the high-$z_h$ hadrons. The reason is again the strong prehadronic interactions of the beam and target remnants right after the $\gamma^*N$ interaction.

When setting the prehadron production time to $0.2\tau_{p2}$ (solid line in Fig.~\ref{fig:emc2}), we observe again a slightly too strong attenuation. However, this concept is in better agreement with the experimental data at EMC energies than our previous approaches.

In summary, it does not seem to be possible to simultaneously describe both the HERMES and the EMC data with (pre-)hadronic FSI only, as long as one does not account for additional effects like color transparency.

\section{Summary and Outlook}
\label{sec:summary}
In this work we have presented a detailed investigation of hadron attenuation in deep-inelastic lepton scattering at HERMES and EMC energies. In the primary production process we do not only consider the direct interaction of the virtual photon $\gamma^*$ with a quark inside the nucleon (nucleus), but also diffractive and hard scatterings of the photon's hadronic components. In the latter case we account for coherence length effects that lead to nuclear shadowing in DIS of nuclei (cf.~Section \ref{sec:shadowing}). Furthermore, our model description incorporates other nuclear effects like Fermi motion of the bound nucleons, Pauli blocking and nuclear binding. We determine the complete final state of each $\gamma^*N$ interaction using the event generator PYTHIA.

For the space-time picture of the hadronization process we basically have considered two different scenarios. One is motivated by the Lund model and the other one is that generally used in standard transport models. In the Lund model the production proper times of the hadron constituents and the formation proper time of the hadron are determined by the underlying fragmentation process and depend on energy and momentum of the hadron. Since PYTHIA is based on the Lund fragmentation scheme we can directly extract the production and formation times for each hadron from the corresponding fragmentation process. We have shown explicitly that the production and formation times in the Lund model show a non-trivial dependence on the mass of the hadron.

On the other hand, the concept generally used in transport approaches is that the production time of the prehadrons is set to zero and the interaction probability is reduced during the formation time $\tau_f$. For simplicity the formation time is assumed to be a constant $\tau_f$ in the rest frame of each hadron in order to work with a single parameter, only.

The prehadronic final state interactions (FSI) between the production and the formation time and the hadronic FSI after $\tau_f$ are described within a transport model which allows for a realistic coupled channel treatment beyond simple absorption mechanisms. We explicitly account for particle creation in the interactions of the primary reaction products emerging from the initial $\gamma^*N$ interaction. These secondary particles are found to strongly influence the low-energy part of the experimentally observed multiplicity ratios (\ref{eq:multiplicity-ratio}).

Furthermore, we have studied how the kinematic cuts and the finite detector acceptance influence the experimental observables. We find strong effects that have to be taken into account in any robust interpretation of the data.

We have investigated the attenuation of charged and neutral pions, kaons, protons and anti-protons as a function of the fractional energy $z_h$, the photon energy $\nu$ and the transverse momentum $p_T$ as well as the double-hadron attenuation. While in the kinematic region of the HERMES experiment most phenomena can be attributed to (pre-)hadronic FSI, we find limitations of our 'standard' model for EMC energies and for large $p_T$. The latter supports a partonic origin for the Cronin effect in electron-nucleus interactions.

We have seen in Sec.~\ref{ssec:HERMES-Lund} that a scenario in which any interactions before the production of the color neutral prehadrons at proper time $\tau_{p2}$ are neglected does not yield enough hadron attenuation. Instead, we had to assume strong interactions right after the production times $\tau_{p1}$ of the hadron's first constituent. However, this does not necessarily imply that there is a freely interacting quark that sees strong FSI. As pointed out in Sec.~\ref{ssec:Lund-prehadrons} the longitudinal dimension of the string and its (color neutral) fragments is not expected to be larger than 1 fm \cite{Gei99,Cio02}. Therefore, one has in principle to deal with the propagation of color neutral strings that subsequently fragment into prehadrons and color neutral remainder strings. A collision of a (remainder) string -- before its fragmentation -- will most likely lead to a different final state than an undisturbed decay. This is technically more involved but will be incorporated in our future work. Apart from extensions on the theoretical side new experimental data also at lower energies should help to clarify the problem of hadronization. The experiments that are currently performed at considerably smaller energies at Jefferson Lab -- and which are planned after the upgrade to 12 GeV beam energy -- will be more sensitive to the string fragmentation and hadron formation times since time dilatation effects are less pronounced.

Furthermore, in a direct photon-nucleon interaction at very high $Q^2$ and energy $\nu$ the photon is expected to knock out a highly virtual point-like quark that immediately may radiate gluons. The gluons then can split into quark-antiquark pairs and finally the various colored quarks and gluons combine to form the hadronic strings. While these processes are in principle taken into account in our present simulations via PYTHIA, we have neglected their space-time evolution so far and assumed that they take place instantaneously at the interaction point. This simplification may have considerable consequences at EMC energies and possibly explain our difficulties in describing the data. An explicit consideration of these partonic evolution effects is also planned for the future.

\begin{acknowledgments}
This work has been supported by BMBF. The authors acknowledge valuable discussions with A.~Accardi, P.~Di Nezza, G.~Elbakyan, A.~Hillenbrand, B.~Z.~Kopeliovich, P.~Liebing, V.~Muccifora, and X.~N.~Wang.
\end{acknowledgments}
\appendix
\section{High-energy hadron-nucleon interactions}
\label{app:hadron-nucleon}
\subsection{Total cross sections}
As pointed out in Sec.~\ref{sec:BUU} we model hadron-hadron collisions above the resonance region ($W\gtrsim2.3$ GeV) by FRITIOF \cite{FRITIOF}. In contrast to PYTHIA \cite{PYTHIA} the FRITIOF event generator does not provide absolute cross sections but only determines the final state of a scattering event. Therefore, we have to explicitly parameterize the total cross sections at large energies. For all baryon-baryon collisions -- that solely involve non-strange baryons in the entrance channel -- we use the cross section parameterization \cite{PDG94}
\begin{equation}
\label{eq:baryon_cs}
    \sigma=\left[A+Bp^n+C\ln^2p+D\ln p\right]\,\mathrm{mb}
\end{equation}
with the parameters for proton-proton collisions listed in Table \ref{tab:baryon_cs}. In Eq.~(\ref{eq:baryon_cs}) $p$ denotes the laboratory momentum in GeV. As shown in Ref.~\cite{PDG94} this parameterization yields a good description of the cross sections down to the resonance region. In the current version of our transport simulation we neglect collisions of strange and charmed baryons.

For the meson-nucleon interactions, that can be addressed experimentally like $\pi^\pm p$, $K^\pm p$ and $K^\pm n$, we use Eq.~(\ref{eq:baryon_cs}) with the parameters listed in Table \ref{tab:baryon_cs}. The cross sections for the other isospin channels are given by isospin symmetry
\begin{eqnarray}
    \sigma_{\pi^0N}&=&\frac{1}{2}(\sigma_{\pi^+p}+\sigma_{\pi^-p})\,,\nonumber\\
    \sigma_{\pi^+n}&=&\sigma_{\pi^-p}\,,\nonumber\\
    \sigma_{\pi^-n}&=&\sigma_{\pi^+p}\,,\nonumber\\
    \sigma_{K^0p}&=&\sigma_{K^+n}\,,\nonumber\\
    \sigma_{\bar{K}^0p}&=&\sigma_{K^-n}\, ,\qquad\mathrm{etc.}
\end{eqnarray}
The cross sections for the $K^*$ and $\bar{K}^*$ mesons are assumed to be the same as for $K$ and $\bar{K}$. The total cross sections for the vector mesons are given by the PYTHIA parameterization (\ref{eq:vector-cs}).

For all other high-energy meson-baryon collisions we employ an ansatz similar to that of the UrQMD model \cite{UrQMD}:
\begin{equation}
    \sigma_{mb}^\mathrm{tot}(s)=\sigma_{\pi N}^\mathrm{tot}(s)
    \big(1-0.4x_S^m-0.5x_C^m\big)\big(1-0.4x_S^b-0.5x_C^b\big) ,
\end{equation}
where $x_{S,C}^m$ and $x_{S,C}^b$ denote the strangeness and charm content of the meson $m$ and baryon $b$, respectively:
\begin{eqnarray}
    x_S^m=\frac{|S(m)|}{2}\, , & &\qquad x_C^m=\frac{|C(m)|}{2}\, ,\nonumber\\
    x_S^b=\frac{|S(b)|}{3}\, , & &\qquad x_C^b=\frac{|C(b)|}{3}\, .
\end{eqnarray}
Here $S$ and $C$ denote the strangeness and charm quantum numbers of the meson and baryon, respectively.

\subsection{Elastic scattering}
\begin{figure}
    \includegraphics[width=13cm]{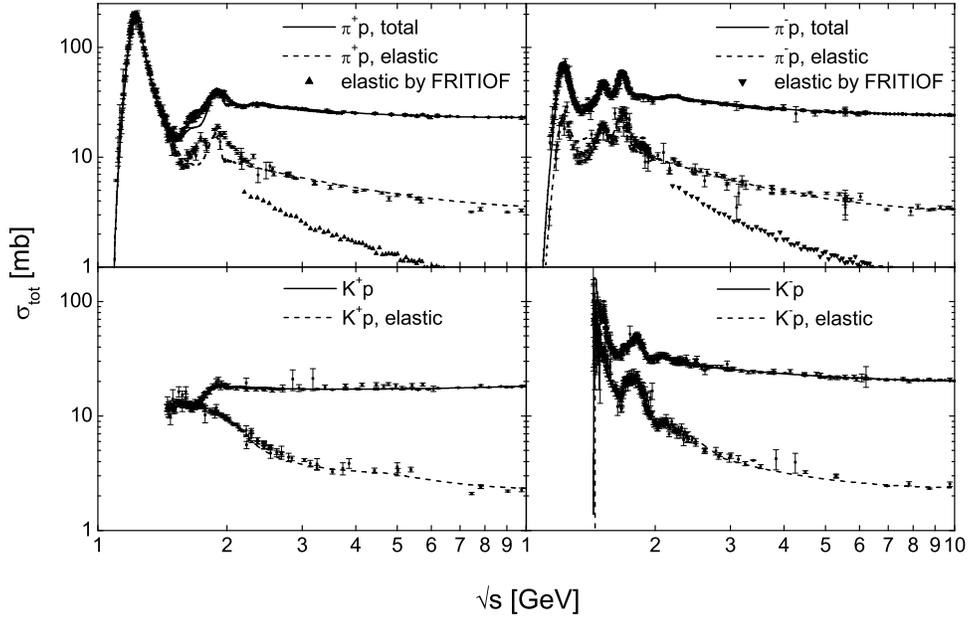}
  \caption{Cross sections for $\pi^\pm p$ and $K^\pm p$ scattering. The solid (dashed) lines represent the total (elastic) cross sections. The high-energy part given by Eq.~(\ref{eq:baryon_cs}) is continuously connected to the cross sections of the resonance model \cite{Eff99c} applied below $\sqrt{s}=2.2$ GeV. The solid triangles show the FRITIOF result for elastic pion-proton scattering. The data are taken from Ref.~\cite{PDG02}.}
\label{fig:cstot}
\end{figure}

There are two problems with the original FRITIOF model \cite{FRITIOF}. The first one is that it does not generate enough elastic scattering events as can be seen by the solid triangles in Fig.~\ref{fig:cstot}. We cure this deficiency by also parameterizing the elastic cross section and simulating elastic scattering externally; FRITIOF is then only called for inelastic collisions (cf.~Ref.~\cite{HSD}). For all elastic baryon-baryon collisions, that solely involve non-strange baryons in the entrance channel, we use the same parameterization as for elastic $pp$ scattering (see Table \ref{tab:baryon_cs}).

The cross sections for elastic $\pi^\pm p$ and $K^\pm p$ scattering are again taken from experiment (see Table \ref{tab:baryon_cs}) and those for elastic $VN$ scattering are given by VMD,
\begin{eqnarray}
    \sigma_{VN}^\mathrm{el}&\approx&\left(\frac{g_V}{e}\right)^2\sigma_{\gamma N\to VN}.\nonumber
\end{eqnarray}
For the remaining particles we use the relation \cite{Gou83},
\begin{equation}
    \label{eq:sigma_el}
    \sigma_\mathrm{el}=c\;\sigma_\mathrm{tot}^{3/2},
\end{equation}
to determine the size of the elastic cross sections from the corresponding total cross sections. The value $c=0.039$ mb$^{-1/2}$ is the same as in the UrQMD model \cite{UrQMD}. The angular distribution for elastic scattering $1+2\to 1+2$ is taken from the PYTHIA model,
\begin{equation}
\label{eq:sigma_el1}
\frac{d\sigma^\mathrm{el}}{dt}\sim e^{B^\mathrm{el}_{1\,2}\,t}\,,
\end{equation}
with the slope parameter
\begin{equation}
    B^\mathrm{el}_{1\,2}=2b_1+2b_2+4s^\epsilon-4.2\nonumber
\end{equation}
with $s$ given in units of GeV$^2$ and $B^\mathrm{el}_{1\,2}$ in GeV$^{-2}$. The constants are $b_b=2.3$ for baryons and $b_m=1.4$ for mesons.

\subsection{Quark-antiquark annihilation}
\begin{figure}[bt]
    \includegraphics[width=12cm]{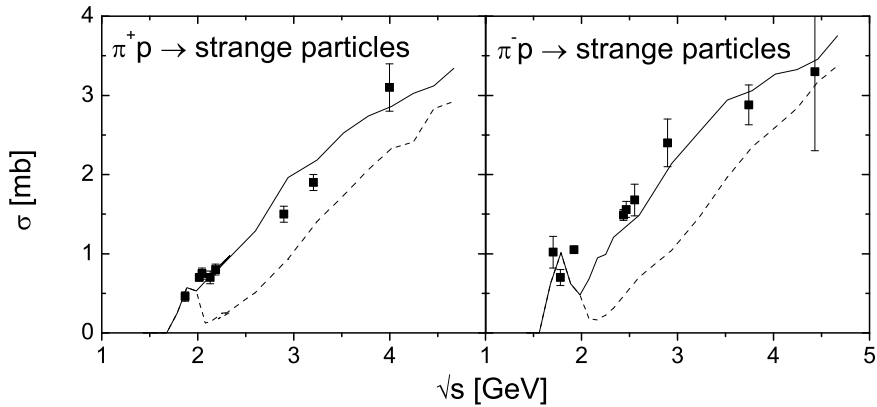}
  \caption{Cross sections for $\pi^\pm p\to$ {\it strange particles} in comparison with experimental data from Ref.~\cite{Lan88}. The solid (dashed) lines represent the simulation with (without) the possibility of quark-antiquark annihilation. The figure is taken from Ref.~\cite{Wag04}.}
\label{fig:manni}
\end{figure}

A second problem with FRITIOF is that it does not account for quark-antiquark annihilation in meson-baryon scattering in correspondence to Reggeon exchange in the $t$ channel. This process, however, gives a finite contribution to the total cross section for low $\sqrt{s}$. We, therefore, keep the option to simulate its contribution to the inelastic collisions independently of FRITIOF using the method developed in Ref.~\cite{Wag04}, where the antiquark from the meson and a constituent of the baryon with the same flavor may annihilate. We assume that the momenta of the two annihilating quarks are very small so that we do not need to treat the final gluon explicitly. The final state of such an annihilation process is modeled by an excited string with the invariant mass of the colliding system. The decay of the string into hadrons is taken into account using the Lund fragmentation routine JETSET 7.3 as also used by FRITIOF. In contrast to the two excited strings of a FRITIOF event, the string -- emerging after $q\bar{q}$ annihilation -- has a larger invariant mass and, therefore, has more energy available for $s\bar{s}$ creation. Hence, the annihilation process has a strong effect on strangeness production in meson-baryon scattering and the energy dependence of its cross section. Above $\sqrt{s}=2.2$ GeV the annihilation cross section relative to strangeness production in pion-nucleon scattering can be fitted by \cite{Wag04}:
\begin{equation}
\frac{\sigma_{mb}^\mathrm{anni}}{\sigma_{mb}^\mathrm{inel}}=\max\left[1.2-0.2\frac{\sqrt{s}}{\mathrm{GeV}}\,,\,0\right].\nonumber
\end{equation}
Fig.~\ref{fig:manni} shows the resulting cross section for strangeness production in $\pi^\pm p$ with (solid line) and without (dashed line) the quark-antiquark annihilation contribution. One clearly sees that the incorporation of the annihilation part leads to a much better agreement with data.

Since the cross sections for baryon-baryon collisions below $\sqrt{s}=2.6$ GeV and meson-baryon collisions below $\sqrt{s}=2.2$ GeV are given by the resonance model of Ref.~\cite{Eff99c}, we continuously connect the high-energy parameterization to the resonance part. As an example, Fig.~\ref{fig:cstot} shows our parameterization of the total and elastic $\pi^\pm p$ and $K^\pm p$ cross section in comparison with the experimental data.

\subsection{Antibaryons}
Finally, there is the possibility of elastic and inelastic baryon-antibaryon ($b\bar{b}$) interactions in the nuclear FSI. The total and elastic cross sections for $p\bar{p}$ scattering are again taken from experiment, i.e.~Eq.~(\ref{eq:baryon_cs}) and Table \ref{tab:baryon_cs}. For the rest of the non-strange antibaryons we use the same cross sections as for antiprotons. Elastic $b\bar{b}$ scattering is simulated in the same way as in all other elastic channels. The inelastic fraction of the total cross section is experimentally known to be dominated by annihilation. We, therefore, reduce the inelastic cross section of (anti-)baryons that involve ($\bar{s}$)$s$ quarks according to the simple valence quark picture:
\begin{equation}
\sigma^\mathrm{anni}_{Y\bar{N}}=\sigma^\mathrm{anni}_{\bar{Y}N}=
\frac{3-|S|}{3}(\sigma_{p\bar{p}}^\mathrm{tot}-\sigma_{p\bar{p}}^\mathrm{el}),
\end{equation}
where $S$ denotes the strangeness of the (anti-)hyperons. The annihilation in collisions involving charmed (anti-)baryons is neglected here. The annihilation process is modeled in the same way as in Ref.~\cite{Gei98}: After annihilation of a quark and an antiquark with the same flavor the remaining (anti-)quarks form two orthogonal $q\bar{q}$ jets which equally share the invariant mass of the colliding system. As in the case of $q\bar{q}$ annihilation in meson-baryon scattering the strings are fragmented using JETSET 7.3.

Throughout this work we neglect meson-meson, meson-antibaryon as well as antibaryon-antibaryon interactions. They are very unlikely in lepton-induced reactions since they require interactions between the few reaction products among each other.

\section{The BUU transport model}
\label{app:BUU-details}
\subsection{Ingredients}
The scalar potential $U_S$ in the set of transport equations (\ref{eq:BUU}) can be related to an effective (non-relativistic) potential $U$ which accounts for the many-body interactions of the baryons among each other. The general expression for the relativistic energy of a particle under the influence of a scalar potential $S$ and a vector potential $V=(V_0,\vec V)$ is 
\begin{equation}
H=\sqrt{(\mu+S)^2+(\vec p-\vec V)^2}+V_0.
\end{equation}
In the local rest frame (LRF), i.e.~where the baryon current locally vanishes, the spatial components $\vec{V}$ vanish. We arbitrarily set $S=0$ and interpret the effective potential $U$ as the zeroth component $V_0$ of the vector potential:
\begin{equation}
    H_\mathrm{LRF}=\sqrt{\mu^2+p_\mathrm{LRF}^2}+U(\vec r,\vec p_\mathrm{LRF}).
\end{equation}
We can then define the scalar potential $U_S$ of Eq.~(\ref{eq:hamilton}) in any frame as
\begin{equation}
    U_S:=\sqrt{H_\mathrm{LRF}^2-p_\mathrm{LRF}^2}-\mu.
\end{equation}
Note that for photon- and electron-induced reactions the local rest frame coincides with the frame where the target nucleus is at rest, i.e.~the laboratory frame in case of fixed target experiments.

For nucleons the effective potential $U$ is parameterized according to Ref.~\cite{Wel88} as a sum of a Skyrme part, which only depends on the baryon density $\rho$, and a momentum dependent part:
\begin{equation}
\label{eq:skyrme-potential}
    U(\vec r,\vec p)=A\frac{\rho(\vec r)}{\rho_0}+B\left(\frac{\rho(\vec r)}{\rho_0}\right)^\tau+\frac{2C}{\rho_0}g\intop\frac{d^3p'}{(2\pi)^3}\frac{f(\vec r,\vec p\,')}{1+\left(\frac{\vec p-\vec p\,'}{\Lambda}\right)^2}
\end{equation}
where $\rho_0=0.168$ fm$^{-3}$ denotes the saturation density of nuclear matter. In the reactions considered in this work the nucleus remains close to its ground state and the phase-space density $f$ in (\ref{eq:skyrme-potential}) can be approximated by the phase space density of (uncorrelated) cold nuclear matter
\begin{equation}
\label{eq:nuclear-matter}
    f(\vec r,\vec p)\sim\Theta(p_F(\vec r)-|\vec p|)
\end{equation}
with the local Fermi momentum
\begin{equation}
    p_F(\vec r)=\left(\frac{6\pi^2}{g}\rho(r)\right)^{1/3}.
\end{equation}
Here $g=4$ again denotes the factor of degeneracy. For the density distribution of complex nuclei we use the Woods-Saxon parameterization:
\begin{equation}
\label{eq:woods-saxon}
    \rho(r)=\frac{\rho_0}{1+\exp\left(\frac{r-R}{a}\right)}
\end{equation}
with the parameters of Table \ref{tab:woods-saxon} that have been extracted from a Hartree-Fock calculation \cite{Len99} for stable nuclei. For light nuclei like $^{14}$N, however, we use a Gaussian shape:
\begin{equation}
\label{eq:gauss}
    \rho_g(r)=\frac{1}{\pi^{3/2}a_g^3}\exp\left(-\frac{r^2}{a_g^2A^{2/3}}\right)
\end{equation}
with $a_gA^{1/3}=\sqrt{\frac{2}{3}}r_\mathrm{rms}$ and $r_\mathrm{rms}=1.21 A^{1/3}$ fm.

The use of (\ref{eq:nuclear-matter}) allows us to employ an analytic expression for the momentum dependent part of the potential (\ref{eq:skyrme-potential})
\begin{eqnarray}
    & &\frac{2C}{\rho_0}g\intop\frac{d^3p'}{(2\pi)^3}\frac{\Theta(p_F(\vec r)-|\vec p\,'|)}{1+\left(\frac{\vec p-\vec p\,'}{\Lambda}\right)^2}=
    \frac{2C}{\rho_0}\frac{g}{(2\pi)^3}\pi\Lambda^3\bigg\{\frac{p_F^2(\vec r)+\Lambda^2-p^2}{2p\Lambda}\ln\left[\frac{(p+p_F(\vec r))^2+\Lambda^2}{(p-p_F(\vec r))^2+\Lambda^2}\right]\nonumber\\
    & &\qquad\qquad+ \frac{2p_F(\vec r)}{\Lambda}-2\left[\arctan\left(\frac{p+p_F(\vec r)}{\Lambda}\right)-\arctan\left(\frac{p-p_F(\vec r)}{\Lambda}\right)\right]\bigg\}.
\end{eqnarray}

The parameters of the mean-field potential (\ref{eq:skyrme-potential}) are fitted to the saturation properties of nuclear matter and the momentum dependence of the nucleon optical potential as measured in $pA$ collisions \cite{Eff99c}. Throughout this work we employ the parameters given in Table \ref{tab:pot-parameter}.

We use the same mean-field potential for all baryons except for the $\Delta$ resonance for which we assume
\begin{equation}
    U_\Delta=\frac{2}{3}U .
\end{equation}
This choice is motivated by the phenomenological value of -30 MeV at density $\rho_0$ \cite{Eri88}. As mentioned before, we neglect any hadronic potential for mesons as well as any influence of the Coulomb potential in our present investigations.

In the multi-GeV range of interest in this work the coupling of the BUU equations (\ref{eq:BUU}) via the mean field is rather low because the absolute value of $U_S$ is about 100 MeV at maximum. One is, therefore, left with the coupling through the collision term (\ref{eq:coll-term}). In case of high-energy lepton-induced reactions binary collisions $a_1,a_2\to b_1,\ldots b_m$ play the dominant role. Since some of the particles in our simulation are unstable with respect to strong decays one has to keep in mind that also the decay of a particle into several other hadrons leads to changes in the phase-space densities.

In the mesonic sector we account for all particles listed in Table \ref{tab:mesons}. Mesons that can only decay due to the weak interaction are considered to be stable within our model. The only exception is the $\eta$ whose decay is explicitly accounted for when evaluating pion production.

Besides the nucleon ($m_N=938$ MeV) and the Delta ($m_\Delta=1232$ MeV, $\Gamma_\Delta=118$ MeV) we account for 29 other nucleon resonances \cite{Eff99c} in the baryonic sector whose properties are taken from an analysis of $\pi N$ scattering \cite{Man92}. In our model the $\Delta$ always decays to $N\pi$ whereas the other resonances can couple to the channels $N\pi$, $N\eta$, $\Lambda K$, $N\omega$, $\Delta\pi$, $N\rho$, $N\sigma$, $N(1440)\pi$ and $\Delta\rho$. In addition to the $\Lambda$ ($m_\Lambda=1116$ MeV) and the $\Sigma$ ($m_\Sigma=1189$ MeV), which are stable with respect to the strong decay, we also include 19 further $S=-1$ resonances that can decay into $\Lambda\pi$, $NK$, $\Sigma\pi$, $\Sigma^*\pi$, $\lambda\eta$, $NK^*$ and $\Lambda^*\pi$. Furthermore, we include the strange and charmed baryons of Table \ref{tab:baryons} in our model. Due to a lack of a complete analysis, the parameters for the strange and charmed baryons are taken from Ref.~\cite{PDG98}. For each baryon we also account for the corresponding antiparticle.

Since we explicitly consider the charge of the particles each isospin state of a particle leads to a separate BUU transport equation. The spin is only accounted for as a statistical weight in the degeneracy factor $g$.

\subsection{Numerical realization}
The set of coupled differential-integral equations (\ref{eq:BUU}) is solved via a test-particle ansatz for the spectral phase-space densities (\ref{eq:spectral-phase-space}):
\begin{equation}
\label{eq:test-particles}
    F(\vec r,\vec p,\mu;t)=\frac{1}{N}\frac{(2\pi)^3}{g}\sum_{i=1}^N\delta(\vec r-\vec r_i(t))\delta(\vec p-\vec p_i(t))\delta(\mu-\mu_i(t))
\end{equation}
where $\vec r_i$, $\vec p_i$ and $\mu_i$ denote the position, momentum and mass of the test particle $i$ at time $t$ and $N$ is the number of test particles per physical particle. In this work we use the method of parallel ensembles, i.e.~the test particles are divided into $N$ different ensembles which do not influence each other. This is equivalent to simulating $N$ independent nuclear reactions in parallel and averaging the observables at the end. For a test-particle number $N \rightarrow \infty$ the test particles will give the time evolution of the spectral phase-space densities.

When initializing a nuclear reaction the test particles, that correspond to nucleons of the nucleus, are distributed in position space following a Woods-Saxon distribution (\ref{eq:woods-saxon}) or a Gauss distribution (\ref{eq:gauss}) for $^{14}$N, respectively. We here assume that the form of the density distribution is the same for protons and neutrons. For the initialization in momentum space we use the local Thomas-Fermi approximation (\ref{eq:nuclear-matter}).

The calculation is performed on a discretized time grid with a default grid size $\Delta t=$ 0.5 fm/$c$. During each time step the test particles are assumed to move as non-interacting particles in the mean field $U_S$. Substituting the test-particle ansatz (\ref{eq:test-particles}) into the BUU equation (\ref{eq:BUU}) -- with the collision term set to zero -- yields the classical Hamilton equations of motion
\begin{eqnarray}
    \frac{d\vec r_i}{dt}&=&\vec \nabla_{\vec p_i}H\nonumber\\
    \frac{d\vec p_i}{dt}&=&-\vec \nabla_{\vec r_i}H\nonumber\\
    \frac{d\mu_i}{dt}&=&0,
\end{eqnarray}
with $H$ being a functional of the phase-space density $f$.

Between the time steps the particles may collide. We do not assume any medium modification of the matrix elements ${\mathcal{M}}$ that enter the collision term (\ref{eq:coll-term}). If one accounts for the energy shift caused by the scalar potential $U_S$, the transition rates can be directly taken from the corresponding vacuum cross sections. Note, however, that in the resonance region cross sections might be modified due to in-medium changes of the resonance properties as discussed in Refs.~\cite{Eff99c,Leh00}. 

Concerning the collision criteria we follow the method by Kodama et al. \cite{Kod84}: Two particles collide in a time step $\Delta t$ if the impact parameter $b$, i.e.~the minimum separation in their center-of-momentum system, is smaller than
\begin{equation} \label{crit}
    b\leq\sqrt{\frac{\sigma^\mathrm{tot}(s)}{\pi}} .
\end{equation}
Furthermore, it is checked if both particles reach this minimal distance during the time step $\Delta t$. In Eq.~(\ref{crit}) $\sigma^\mathrm{tot}$ denotes the total cross section for the interaction of the two particles. For high-energy collisions these are the ones given in Appendix \ref{app:hadron-nucleon}. Elastic interactions occur with a probability
\begin{equation}
    P_\mathrm{el}=\frac{\sigma^\mathrm{el}(s)}{\sigma^\mathrm{tot}(s)}
    ;
\end{equation}
 the scattering angle is determined according to Eq.~(\ref{eq:sigma_el1}). In case of a high energy inelastic collision the reaction products are determined by FRITIOF (or JETSET for baryon-antibaryon annihilation). In the resonance region, i.e.~below $\sqrt{s}=2.2$ GeV for meson-baryon and $\sqrt{s}=2.6$ for baryon-baryon scattering, the total cross section is an incoherent sum of the cross sections for the reactions
\begin{eqnarray}
    mB \leftrightarrow R,\quad \pi N \leftrightarrow \pi N,\quad \pi N \rightarrow \pi\pi N,\quad \pi N \leftrightarrow \eta\Delta,\quad \pi N \leftrightarrow \omega N,\quad \pi N \rightarrow \pi\omega N,\nonumber\\
     \pi N \leftrightarrow \phi N,\quad \omega N \leftrightarrow \omega N,\quad \omega N \rightarrow \pi\pi N,\quad \phi N \leftrightarrow\phi N,\quad \phi N \rightarrow \pi\pi N,\quad \pi B\leftrightarrow KY,\nonumber\\
     \pi B\rightarrow K\bar{K}N,\quad \bar{K}N\leftrightarrow \bar{K}N,\quad
    \bar{K}N \leftrightarrow \pi Y,\quad \bar{K}N\leftrightarrow \pi Y^*,\quad KN\leftrightarrow KN,\quad KN\rightarrow K\pi N\nonumber
\end{eqnarray}
in case of meson-baryon collisions and
\begin{eqnarray}
    NN\leftrightarrow NN,\quad NN\leftrightarrow NR,\quad NN\leftrightarrow \Delta\Delta,\quad NN\leftrightarrow NN\pi,\quad NN\rightarrow NN\omega,\nonumber\\
     NR\leftrightarrow NR',\quad BB\rightarrow NYK,\quad BB\rightarrow NY^*K,\quad BB\rightarrow NNK\bar{K}\nonumber
\end{eqnarray}
for baryon-baryon collisions. Here $m$ stands for a meson, $B$=$N,\Delta$; nucleon resonances are denoted by $R$ and $R'$, hyperon resonances by $Y^*$ and $Y$=$\Sigma,\Lambda$. The reactions involving antibaryons are obtained by charge conjugation. The reaction channel $ab\to f$ in the collision of two particles $a$ and $b$ is chosen by Monte Carlo with a probability determined from its contribution to the summed total cross section (see Ref.~\cite{Eff99c} for details)
\begin{equation}
\label{eq:monte-carlo}
    P_{ab\to f}=\frac{\sigma_{ab\to f}(s)}{\sigma_{ab}^\mathrm{tot}(s)}.
\end{equation}

An important feature of our model is the decay of unstable particles with mass $\mu$ and energy $E$ during a time step $\Delta t$. The corresponding decay probability is given by 
\begin{equation}
    P_\mathrm{dec}=1-\exp\left(-\frac{\Gamma(\mu)}{\gamma}\Delta
    t\right),
\end{equation}
where $\gamma=E/\mu$ is the Lorentz factor while $\Gamma$ denotes the width of the particle in its rest frame. The final state of the decay is again determined by Monte Carlo assuming the decay to be isotropic in the rest frame of the particle since we neglect the spin degree of freedom.

Due to the low densities of other baryons Pauli blocking is only accounted for in collisions and decays that involve nucleons in the final state. For lepton-induced reactions, where the nucleus approximately stays in its ground state, one can approximate the probability that an event with an outgoing nucleon of momentum $\vec p$ is Pauli blocked via Eq.~(\ref{eq:nuclear-matter}) as
\begin{equation}
    P_\mathrm{Pauli}=\Theta(p_F(\vec r)-|\vec p|).
\end{equation}


\clearpage

\begin{table}[tb]
\vspace{2mm}
\begin{center}
    \begin{tabular}{|c||c|c|c|c|c|}
        \hline
         & $A$ & $B$ & $n$ & $C$ & $D$\\
        \hline
        $\sigma_{pp}^\mathrm{tot}$ & 48.0 & 0. & -- & 0.522 & -4.51\\\hline
        $\sigma_{pp}^\mathrm{el}$ & 11.9 & 26.9 & -1.12 & 0.169 & -1.85\\\hline
        $\sigma_{\bar{p}p}^\mathrm{tot}$ & 38.4 & 77.6 & -0.64 & 0.26 & -1.2\\\hline
        $\sigma_{\bar{p}p}^\mathrm{el}$ & 10.2 & 52.7 & -1.16 & 0.125 & -1.28\\\hline
        $\sigma_{\pi^+p}^\mathrm{tot}$ & 16.4 & 19.3 & -0.42 & 0.19 & 0.\\\hline
        $\sigma_{\pi^+p}^\mathrm{el}$ & 0. & 11.4 & -0.4 & 0.079 & 0.\\\hline
        $\sigma_{\pi^-p}^\mathrm{tot}$ & 33.0 & 14.0 & -1.36 & 0.456 & -4.03\\\hline
        $\sigma_{\pi^-p}^\mathrm{el}$ & 1.76 & 14.0 & -1.36 & 0.456 & -4.03\\\hline
        $\sigma_{K^+p}^\mathrm{tot}$ & 18.1 & 0. & -- & 0.26 & -1.\\\hline
        $\sigma_{K^+p}^\mathrm{el}$ & 5. & 8.1 & -1.8 & 0.16 & -1.3\\\hline
        $\sigma_{K^+n}^\mathrm{tot}$ & 18.7 & 0. & -- & 0.21 & -1.3\\\hline
        $\sigma_{K^-p}^\mathrm{tot}$ & 32.1 & 0. & -- & 0.66 & -5.6\\\hline
        $\sigma_{K^-p}^\mathrm{el}$ & 7.3 & 0. & -- & 0.29 & -2.4\\\hline
        $\sigma_{K^-n}^\mathrm{tot}$ & 25.2 & 0. & -- & 0.38 & -2.9\\
        \hline
    \end{tabular}
\end{center}
\caption{Parameters for the cross section parameterization (\ref{eq:baryon_cs}). The parameters are taken from Ref.~\cite{PDG94}.}
\label{tab:baryon_cs}
\end{table}

\begin{table}[tb]
\vspace{2mm}
\begin{center}
    \begin{tabular}{|c|c|c|c|c|}
        \hline
        $A$ [MeV]& $B$ [MeV] & $C$ [MeV] & $\tau$ & $\Lambda$ [fm$^{-1}$]\\
        \hline
        $-29.3$ & 57.2 & -63.5 & 1.76 & 2.13\\
        \hline
    \end{tabular}
\end{center}
\caption{Parameters of the nucleon potential (\ref{eq:skyrme-potential}) used in this work.}
\label{tab:pot-parameter}
\end{table}

\begin{table}[tb]
\vspace{2mm}
\begin{center}
    \begin{tabular}{|c||c|c|c|c|c|c|l|}
        \hline
& $m$   & $\Gamma$  & & & & &\\
                                    &   [MeV]   &   [MeV]           &   \raisebox{1.5ex}[0cm][0cm]{$J$} &   \raisebox{1.5ex}[0cm][0cm]{$I$}     &   \raisebox{1.5ex}[0cm][0cm]{$S$}     &   \raisebox{1.5ex}[0cm][0cm]{$C$}     & \raisebox{1.5ex}[0cm][0cm]{decay channel}\\
        \hline
        $\pi$           & 138 & 0   & 0 & 1 & 0 & 0 &                                     \\
        \hline
        $ \eta $    & 547   &   $1.2\cdot10^{-3}$   &   0   &   0   &   0   &   0   &   $\gamma\gamma$ (40\%), $\pi^+\pi^-\pi^0$ (28\%), $3\pi^0$ (32\%)\\
        \hline
        $\rho$      & 770 & 151 &   1   &   1   &   0   &   0   &   $\pi\pi$\\
        \hline
        $\sigma$    &   800 &   800 &   0   &   0   &   0   &   0   &   $\pi\pi$\\
        \hline
        $\omega$    &   782 &   8.4 &   1   &   0   &   0   &   0   &   $\pi\pi$ (2\%), $\pi^0\gamma$ (9\%), $\pi^+\pi^-\pi^0$ (89\%)\\
        \hline
        $\eta'$     &   958 &   0.2 &   0   &   0   &   0 & 0   &   $\rho^0\gamma$ (31\%), $\pi\pi\eta$ (69\%)\\
        \hline
        $\phi$      &   1020&   4.4 &   1   &   0   &   0   &   0   &   $\rho\pi$ (13\%), $K\bar{K}$ (84\%), $\pi^+\pi^-\pi^0$ (3\%)\\
        \hline
        $K$             &   496 &   0       &   0   &1/2&   1   &   0   &   \\
        \hline
        $\bar{K}$   &   496 &   0       &   0   &1/2&-1 &   0   & \\
        \hline
        $K^*$           &   892 &   50  &   1   &1/2&   1   &   0   &   $K\pi$\\
        \hline
        $\bar{K}^*$&892 &   50  &   1   &1/2&-1 &   0   &   $\bar{K}\pi$\\
        \hline
        $\eta_c$    &   2980&   0       &   0   &   0   &   0   &   0   &   \\
        \hline
        $J/\psi$    &   3097&   0       &   1   &   0   &   0   &   0   &   \\
        \hline
        $D$             &   1869&   0       &   0   &1/2&   0   &   1   &   \\
        \hline
        $\bar{D}$   &   1869&   0       &   0   &1/2&   0   &-1 & \\
        \hline
        $D^*$           &   2007&   1       &   1   &1/2&   0   &   1   &   $D\pi$\\
        \hline
        $\bar{D}^*$&2007&   1       &   1   &1/2&   0   &-1 &   $\bar{D}\pi$\\
        \hline
        $D_s$           &   1969&   0       &   0   &   0   &   1   &   1   &   \\
        \hline
        $\bar{D}_s$&1969&   0       &   0   &   0   &   -1& -1& \\
        \hline
        $D^*_s$     &   2112&   1.9 &   1   &   0   &   1   &   1   & $D_s\gamma$ (94\%), $D_s\pi$ (4\%)\\
        \hline
        $\bar{D}^*_s$&2112&1.9& 1   &   0   &   -1& -1& $\bar{D}_s\gamma$ (94\%), $\bar{D}_s\pi$ (4\%)\\
        \hline
    \end{tabular}
\end{center}
\caption{List of mesons included in the BUU model, their quantum numbers (spin $J$, isospin $I$, strangeness $S$, charm $C$) and decay channels.~The parameters are taken from Ref.~\cite{PDG98} except for the $\sigma$ meson \cite{Man92}.} \label{tab:mesons}
\end{table}

\begin{table}[tb]
\vspace{2mm}
\begin{center}
    \begin{tabular}{|c||c|c|c|c|c|c|l|}
        \hline
         & $m$  & $\Gamma$  & & & & &\\
                                    &   [MeV]   &   [MeV]           &   \raisebox{1.5ex}[0cm][0cm]{$J$} &   \raisebox{1.5ex}[0cm][0cm]{$I$}     &   \raisebox{1.5ex}[0cm][0cm]{$S$}     &   \raisebox{1.5ex}[0cm][0cm]{$C$}     & \raisebox{1.5ex}[0cm][0cm]{decay channel}\\
        \hline
        $\Xi$                   & 1315  & 0     &   1/2 &   1/2 & -2    &   0   &   \\
        \hline
        $\Xi^*$             &   1530    &   9.5 &   3/2 &   1/2 &   -2  &   0   &   $\Xi\pi$\\
        \hline
        $\Omega$            &   1672    &   0       &   3/2 &   0       &   -3  &   0   &   \\
        \hline
        $\Lambda_c$     &   2285    &   0       &   1/2 &   0       &   0       &   1   &   \\
        \hline
        $\Sigma_c$      &   2455    &   0       &   1/2 &   1       &   0       &   1   &   \\
        \hline
        $\Sigma^*_c$    &   2530    &   15  &   3/2 &   1       &   0       &   1   &   $\Lambda_c\pi$\\
        \hline
        $\Xi_c$             &   2466    &   0       &   1/2 &   1/2 &   -1  &   1   &   \\
        \hline
        $\Xi^*_c$           &   2645    &   4       &   3/2 &   1/2 &   -1  &   1   &   $\Xi_c\pi$\\
        \hline
        $\Omega_c$      &   2704    &   0       &   1/2 & 0  &  -2  &   1   & \\
        \hline
    \end{tabular}
\end{center}
\caption{List of baryons with $S<-1$ or $C>0$ included in the BUU model, their quantum numbers (spin $J$, isospin $I$, strangeness $S$, charm $C$) and decay channels. The parameters are taken from Ref.~\cite{PDG98}.}
\label{tab:baryons}
\end{table}

\begin{table}[tb]
\vspace{2mm}
\begin{center}
    \begin{tabular}{|c||c|c|c|}
        \hline
        $ $ & $R$ [fm] & $a$ [fm] & $\rho_0$ [fm$^{-3}$]\\
        \hline
        $^{14}$N & 2.476 & 0.479 & 0.161\\\hline
        $^{20}$Ne & 2.851 & 0.479 & 0.161\\\hline
        $^{63}$Cu & 4.409 & 0.477 & 0.157\\\hline
        $^{84}$Kr & 4.911 & 0.476 & 0.155\\\hline
        $^{131}$Xe& 5.777 & 0.476 & 0.152\\
        \hline
    \end{tabular}
\end{center}
\caption{Woods-Saxon parameters (\ref{eq:woods-saxon}) for the nuclei investigated in this work.} \label{tab:woods-saxon}
\end{table}

\clearpage

\end{document}